\documentclass[usenatbib]{mn2e}
\usepackage{graphicx}
\usepackage{amsmath}
\usepackage{amssymb}

 \usepackage{hyperref}
 
 \providecommand{\adsurl}[1]{\href{#1}{ADS}}
  
\newcommand{\be}{\begin{equation}}

\newcommand{\ee}{\end{equation}}
\newcommand{\ba}{\begin{eqnarray}}
\newcommand{\ea}{\end{eqnarray}}
\newcommand\bp{\begin{figure}}
\newcommand\ep{\end{figure}}
\newcommand\bpm{\begin{figure*}}
\newcommand\epm{\end{figure*}}
\newcommand{\btab}{\begin{tabular}}
\newcommand{\etab}{\end{tabular}}
\newcommand{\bt}{\begin{table}}
\newcommand{\et}{\end{table}}
\newcommand{\ben}{\begin{enumerate}}
\newcommand{\een}{\end{enumerate}}

\newcommand{\gsim}{ \mathop{}_{\textstyle \sim}^{\textstyle >} }
\newcommand{\lsim}{ \mathop{}_{\textstyle \sim}^{\textstyle <}}
\newcommand{\Fermi}{\emph{Fermi }}

\newcommand{\bcn}{\begin{center}}
\newcommand{\ecn}{\end{center}}

\begin{document}

\title{A statistical test of emission from unresolved point sources}

\author[Tracy R. Slatyer and Douglas P. Finkbeiner]
{Tracy~R.~Slatyer,$^{1,2}$
Douglas~P.~Finkbeiner,$^{1,2}$ \\
$^1$ Physics Department, Harvard University,  Cambridge, MA 02138 USA\\
$^2$ Institute for Theory and Computation, Harvard-Smithsonian Center for Astrophysics, 60 Garden Street, Cambridge, MA 02138 USA}

\maketitle

%------------------------------------------------------------------------------
% SECTION: Abstract
%------------------------------------------------------------------------------
\begin{abstract}
We describe a simple test of the spatial uniformity of an ensemble of discrete events.  Given an estimate for the point source luminosity function and an instrumental point spread function (PSF), a robust upper bound on the fractional point source contribution to a diffuse signal can be found.  We verify with Monte Carlo tests that the statistic has advantages over the two-point correlation function for this purpose, and derive analytic estimates of the statistic's mean and variance as a function of the point source contribution.  As a case study, we apply this statistic to recent gamma-ray data from the \Fermi Large Area Telescope (LAT), and demonstrate that at energies above 10 GeV, the contribution of unresolved point sources to the diffuse emission is small in the region relevant for study of the WMAP Haze.  \end{abstract}


%------------------------------------------------------------------------------
% SECTION: Introduction
%------------------------------------------------------------------------------
\section{Introduction}

Statistical tests of isotropy have a long history in astronomy.  A
common question is ``What fraction of the observed emission could
originate from unresolved point sources?''  For example, possible point source
contributions to the extragalactic X-ray background were investigated
by \cite{1974MNRAS.166..329S}, and the small-angle power spectrum of the
cosmic far infra-red background has been used to estimate the
isotropic component \citep{1996ApJ...473L...9K}.  More recently the
Auger team tested the isotropy of ultra-high energy cosmic ray events
by cross correlating with positions of known active galactic nuclei (AGN;
\cite{Cronin:2007zz,Abraham:2007si}) to provide information on their
origin.  However in the absence of an appropriate external catalog, such 
cross-correlation methods cannot be used, motivating consideration of a 
more general approach. 

In some cases a detector
provides binned counts (e.g. pixels in a CCD); in other cases, photon
event directions are reconstructed in some other way (e.g. a gamma-ray
pair conversion telescope).  In the latter case, it is desirable to apply
statistics that do not require binning of the data, as binning
introduces additional arbitrary parameters into the problem.  In the
limit of low flux density, where the mean density of photon events
(hereafter, ``events'') is much less than one per PSF, explicit
detection of point sources may become impractical, and estimation of the
unresolved point source flux becomes especially difficult.  In some
cases, the two-point correlation function, or some modified form
\citep[e.g.][]{Ave:2009id}, is used as a test.  However, the Fourier
transform of a field of point sources has significant phase correlation,
and a two-point function (or a power spectrum) discards this phase
information.  Higher order correlation statistics capture it, but are
somewhat cumbersome to use.  In the following, we describe a statistic
that is easy to understand and evaluate, and is optimized to address
this question, particularly in the case of fairly sparse data sets with
(on average) $\lesssim 1$ event per PSF circle.

The key insight is that if a substantial fraction of the photons come
from point sources, it is much more likely that two photons appear within
one PSF of each other than in the diffuse case.  This is true even if the
expected integrated flux is of order one count.  Likewise, the number of
PSF circles containing \emph{no} counts is larger if point sources
contribute.  These considerations motivate us to define a ratio between
the fraction of ``isolated events'' and the fraction of ``empty
circles.''  This ratio is very closely related to the fraction of
diffuse emission, and can be calibrated with Monte Carlo simulations for
specific choices of instrumental parameters and a putative luminosity
function.  The two-point function, in contrast, is weighted by the
density squared and is \emph{not} proportional to the desired quantity.

In the following sections we define the statistic, estimate its
variance, show how it behaves in various limits, generalize it to the
case where many events appear in every PSF circle, and show a practical
application to recently released data from the \Fermi Gamma-ray Space
Telescope.

\begin{figure*}
\includegraphics[width=0.48\textwidth]{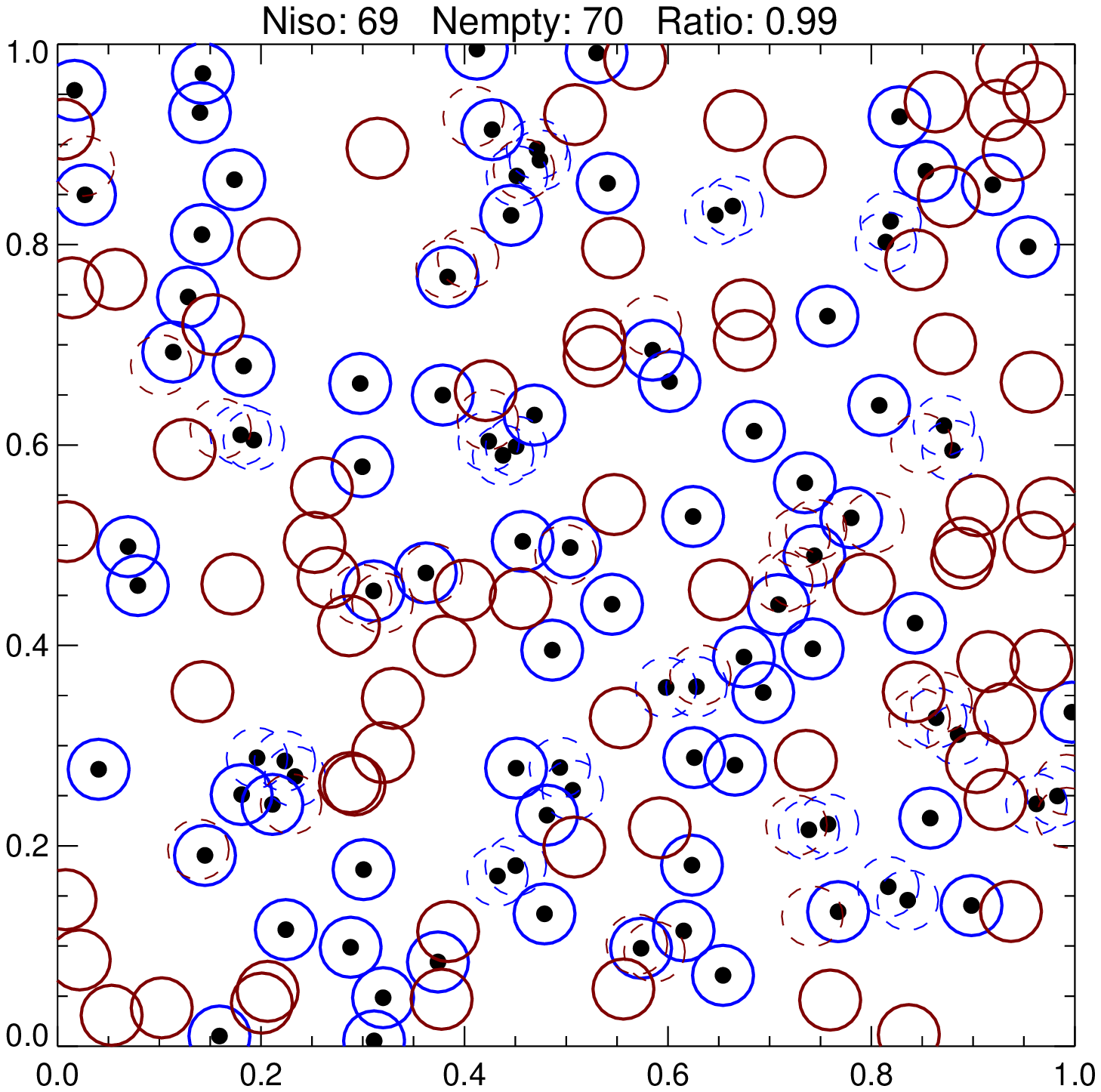}
\includegraphics[width=0.48\textwidth]{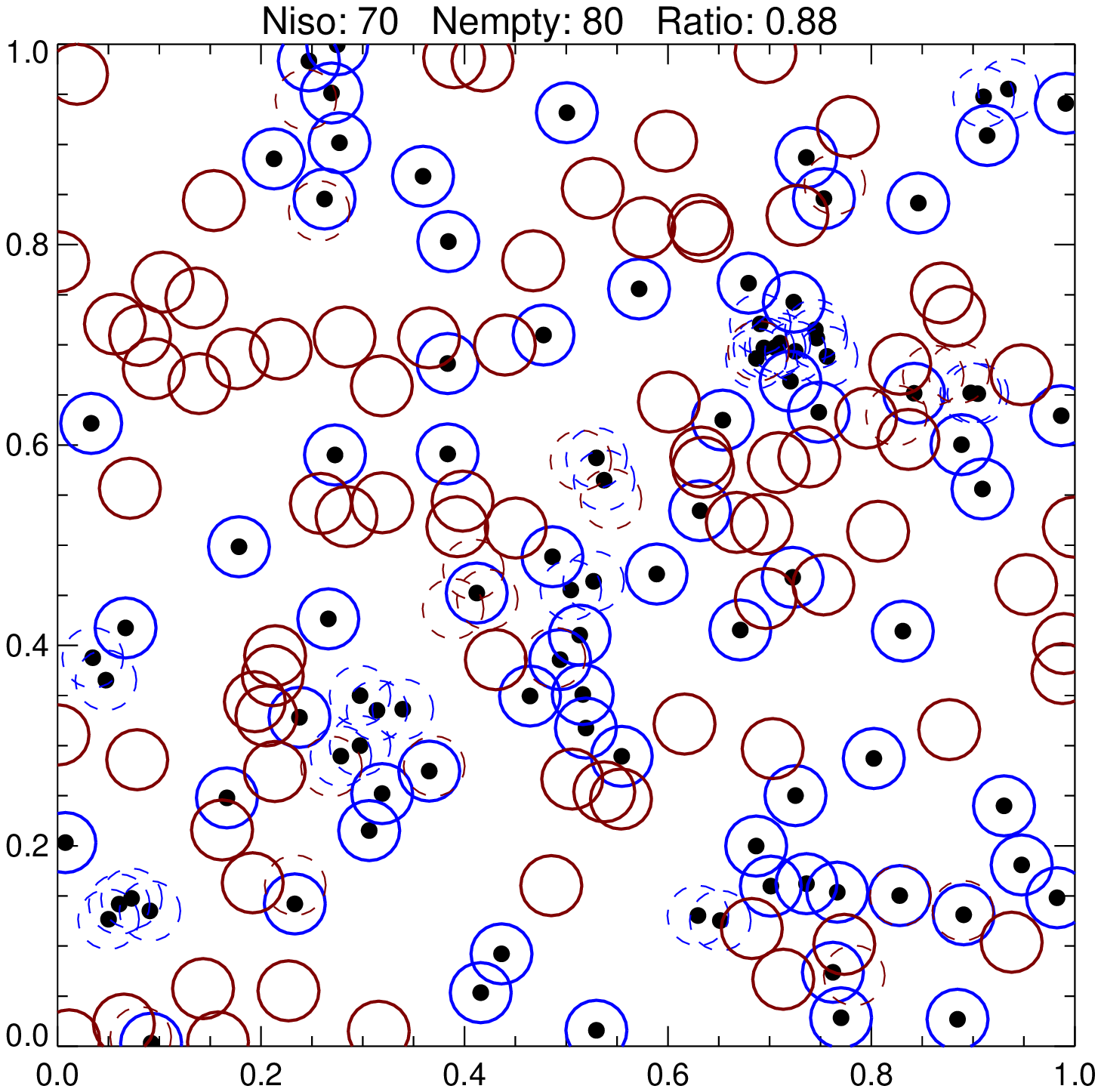}
\includegraphics[width=0.48\textwidth]{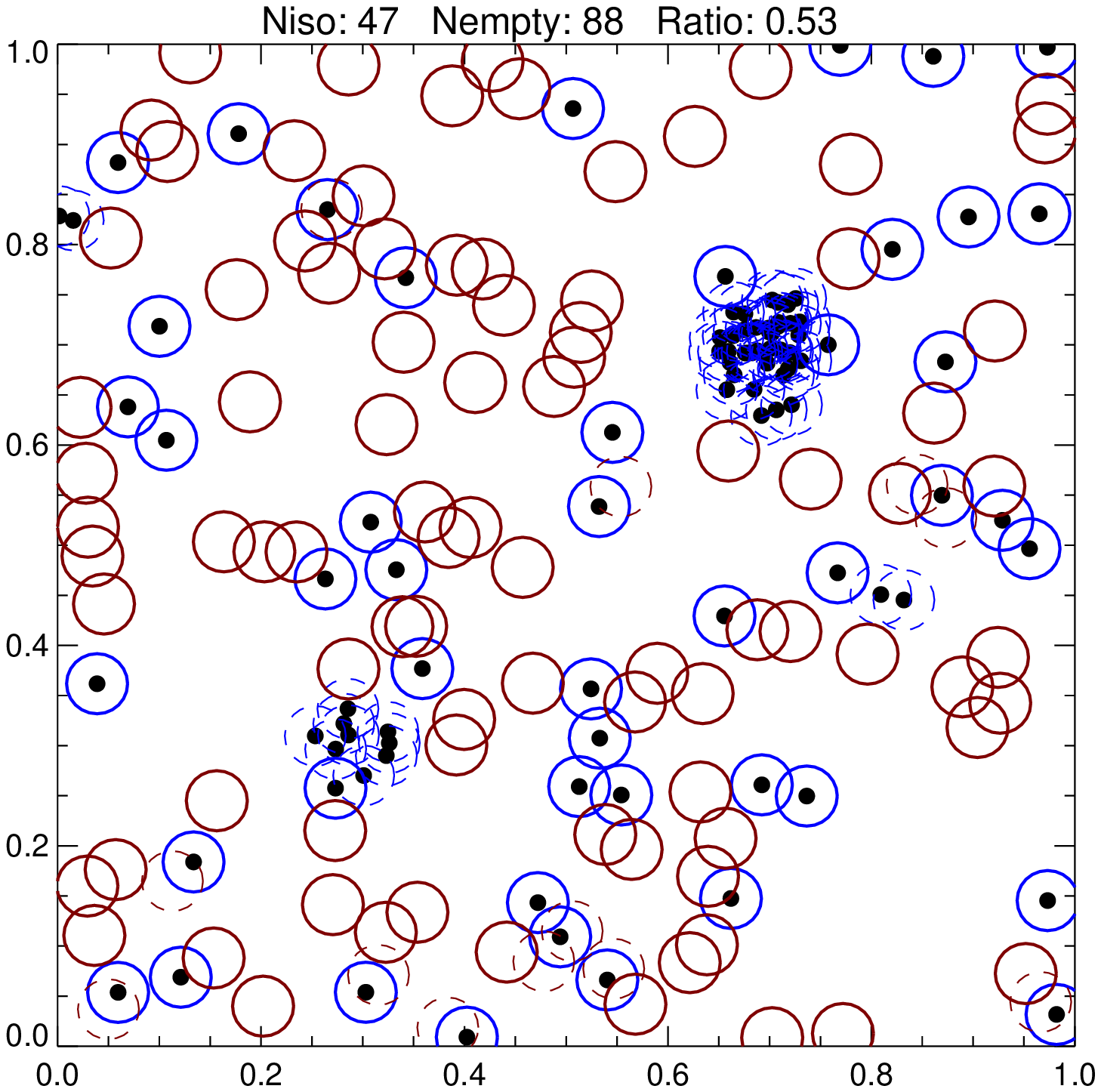}
\includegraphics[width=0.48\textwidth]{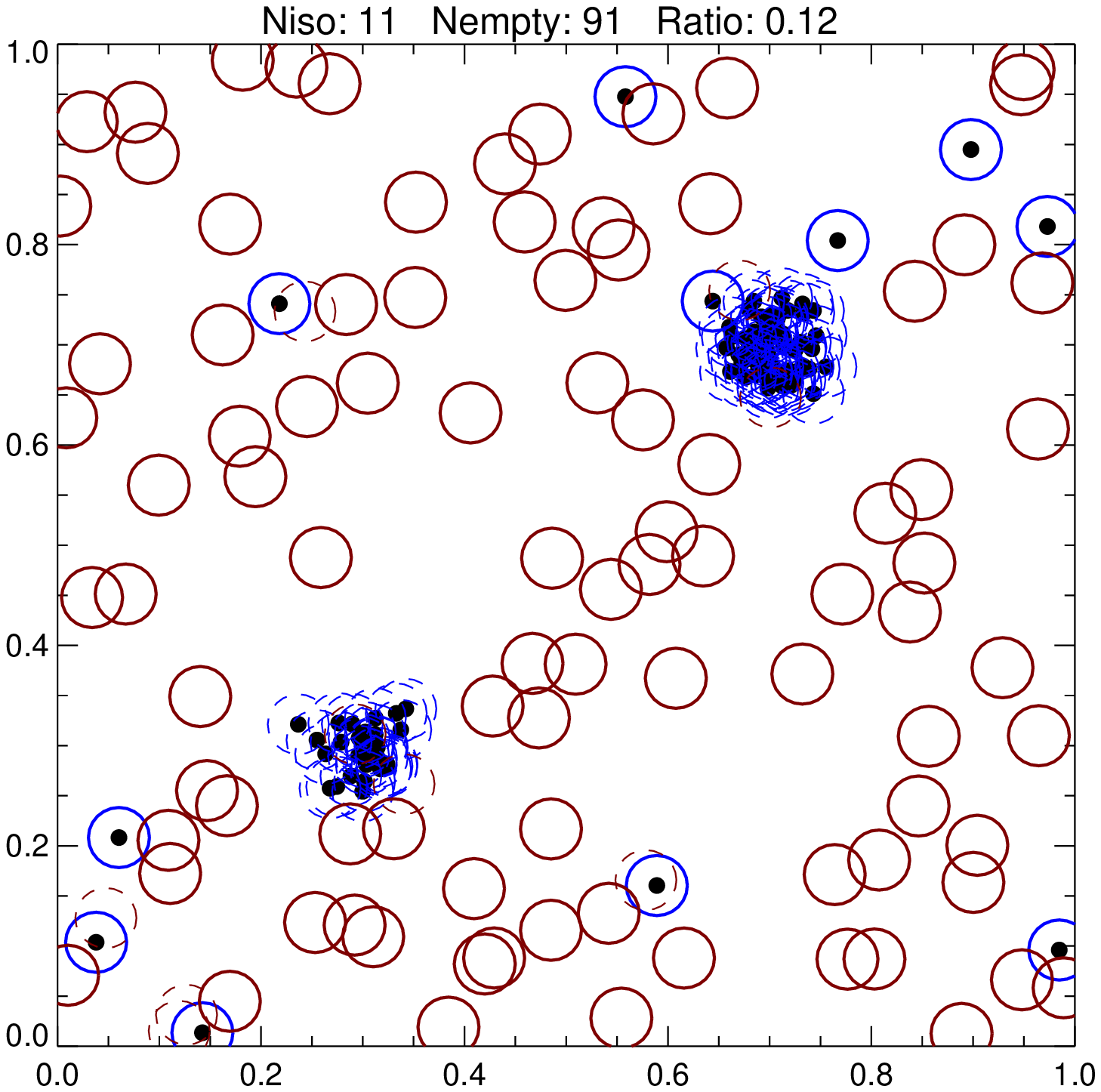}
\caption{\label{fig:circles} 
In each panel, photon events (\emph{dots}) are either isolated
(\emph{solid blue circles}) or not (\emph{dashed blue circles}).  Random
circles are either empty (\emph{solid red}) or not(\emph{dashed red}).
In each case, 100 events and 100 random circles are shown, so the
ratio, $R$, can be visualized here as the number of solid blue circles
divided by the number of solid red circles.  In practise, one uses a
large number of random circles to reduce noise.  The panels contain
either no point sources (\emph{upper left}), or 15\% (\emph{upper
right}), 50\% (\emph{lower left}), or 90\% (\emph{lower right}) point
source flux in sources located at (0.3,0.3) and (0.7,0.7).
}
\end{figure*}

\section{Definition of the statistic}

For each event, we consider the number of neighbouring events within some
test radius $r$; the natural choice of $r$ is determined by the PSF of
the detector. Let the fraction of events with zero neighbours (``isolated
events'') be denoted $n_I$. Now consider a random distribution of points
within the signal region, and consider the number of events within $r$
of each of these points. Let the fraction of points with zero events
within radius $r$ (``empty circles'') be denoted $n_E$ (the number of randomly distributed points should be as large as is computationally feasible, to reduce Poisson error in the fraction $n_E$).

Fig. \ref{fig:circles} illustrates this idea.
If the signal photons are randomly distributed, as in the case of a
uniform diffuse signal, $n_I \approx n_E$ (up to Poisson
fluctuations). If the events are clustered on the scale $r$, on the
other hand, $n_I$ falls (as more events have close neighbours) and
$n_E$ rises (as the events are clumped, more of the signal region
contains no events at all).

If a large fraction of the photon counts originate from unresolved point sources, we expect significant correlations between photon positions, even if the expected counts from a point source are $\lesssim 1$. Consequently, the ratio $R = n_I/n_E$ serves as a simple measure of the fraction of flux originating from point sources as opposed to uniform diffuse emission. 

Larger-scale non-uniformities, such as a gradient in the distribution of diffuse photons, have only a small effect on $R$, as we show explicitly with Monte Carlo tests in \S \ref{sec:results}. In regions of high event density both $n_I$ and $n_E$ are suppressed, so the suppression largely cancels out in the ratio. However, such a gradient \emph{will} tend to slightly lower $R$ in the diffuse limit, since the points relevant to computing $n_E$ are uniformly distributed, whereas those relevant to computing $n_I$ are concentrated in regions of high flux (this in turn can reduce the gradient of $R$ with respect to the fraction of diffuse flux, since in the point-source-dominated limit $R$ is small but independent of any inhomogeneity in the diffuse emission). The point source emission derived from $R$ is thus best interpreted as an upper bound, in cases where the diffuse emission is suspected to have significant spatial variation.

\section{Analytic estimate for the mean and variance}
\label{sec:analytic}

To understand the behaviour of the ratio $R$ as a function of the diffuse emission fraction, consider a related but simpler problem, where we
treat the signal region as a grid and count the number of events in each
cell. In this case an ``isolated'' event is one with no other
events in the same cell, and $R$ is simply the ratio of the
fraction of events which are isolated to the fraction of cells
which are empty.  The total number of cells is $N$, $n$ of which are
empty and $m$ of which have a single event. 

The probability of any given cell being empty is $p_0$, and of
containing a single event, $p_1$.  That is, $\langle n\rangle = p_0N$,
and $\langle m\rangle = p_1N$.  The joint probability of having
precisely $n$ empty cells and $m$ single-event cells is
\begin{equation} P(m,n) = p_0^n p_1^m (1-p_0 - p_1)^{N - n - m}
  \frac{N!}{n! m! (N - m - n)!}.
\end{equation}

The fraction of isolated events is $\langle n_I\rangle =
m/N_\mathrm{event}$, while the empty fraction is $\langle n_E\rangle =
n/N$.  In this case $R = m / (n+1) \times$
$N/N_\mathrm{event}$. As we shall see, $\langle m/(n+1)\rangle$ is an
unbiased estimator of the probability ratio $p_1/p_0$, which corresponds
to $R$ as defined above in the limit of large $n$, and is equivalent to
the fraction of the flux originating from diffuse emission if the point
sources are sufficiently bright.

The expectation value of $m/(n+1)$ is given by,
\begin{equation} \left\langle \frac{m}{n+1}\right\rangle = \sum_{n=0}^{N-1} \sum_{m=1}^{N-n} \frac{m}{n+1} P(m,n) = \frac{p_1 }{p_0} \left(1 -(1-p_0)^N \right). \end{equation}
The expectation value $\langle (m/(n+1))^2 \rangle$ is given by,
\begin{eqnarray}& &  \sum_{n=0}^{N-1} \sum_{m=1}^{N-n} \left(\frac{m}{n+1}\right)^2 P(m,n) \nonumber \\ & = & \frac{p_1}{(-1+p_0)^2 p_0}  \left(-\left(-1+(1-p_0)^N \right) (-1+p_0) p_1 \right. \nonumber \\ & & \left. + N (1-p_0)^N p_0 (1-p_0+N p_1) \right. \nonumber \\ & & \left. \times \,_3 F_2 \left[\{1,1,1-N\},\{2,2\},\frac{p_0}{-1+p_0}\right]\right) \nonumber \\ & \rightarrow & \left( \frac{p_1}{p_0} \right)^2 + \frac{p_1}{N p_0^2} \left(1 + \frac{p_1}{p_0}\right), \quad N \rightarrow \infty. \end{eqnarray}
Consequently, neglecting terms suppressed by a large power of $(1 - p_0)$, the mean $\mu$ and variance $\sigma^2$ of $m/(n+1)$ in the case of large $N$ are given by,
\begin{equation} \mu = p_1/p_0, \quad \sigma^2 = p_1 (1 + p_1/p_0) / N
  p_0^2. 
\end{equation}
The corresponding quantities for $R$ are obtained by a rescaling by
$N/N_\mathrm{event}$ and ($N/N_\mathrm{event})^2$, respectively:
\begin{equation} 
\langle R \rangle = \frac{p_1}{p_0}\frac{N}{N_\mathrm{event}}, \quad 
\left(\frac{\sigma_R}{\langle R \rangle}\right)^2 =  \frac{1 +
  p_1/p_0}{N p_1}
\sim \frac{1}{m}. \end{equation}
It is unsurprising that the fractional uncertainty in $R$ is approximately
$1/\sqrt{m}$ where $m$ is the number of isolated events.  

In the limit where all the flux is diffuse with mean rate
$\lambda$, then $p_1 = \lambda
e^{-\lambda}$, $p_0 = e^{-\lambda}$, and $\langle N_\mathrm{event}\rangle = N \lambda$, so $R$ has mean 1 and variance $(1 + \lambda) e^\lambda /N \lambda$. Now consider the addition of point sources. We can approximate the effect of adding point sources by choosing $T$ cells which each gain $>1$ events. Let the total number of added events be $\Gamma$. Then on average, a fraction $T/N$ of the isolated events will no longer be isolated, and a fraction $T/N$ of the empty cells will no longer be empty (by assuming each cell that gains events gains $>1$, we ensure that empty cells will not gain isolated events).

Thus when point sources are added, both $p_0$ and $p_1$ are multiplied
by $(1 - T/N)$, and $N_\mathrm{event}$ becomes $N \lambda + \Gamma$. Thus the mean of $R$ becomes $N \lambda / (N \lambda + \Gamma)$ = diffuse flux / total flux, and the variance becomes,
\begin{eqnarray} \sigma^2(R) &  = & \frac{1}{N \lambda + \Gamma} \left( \frac{\left(1 + \lambda\right) e^{\lambda}}{(1 + \Gamma / N \lambda) (1 - T/N)} \right) \\
& = & \frac{\left(1 + \lambda \right) e^\lambda}{\mathrm{total} \, \mathrm{counts}} \nonumber \\ & & \times \left( \frac{\text{diffuse flux} / \text{total flux}}{\text{fraction of pixels with no point sources}} \right). \nonumber \end{eqnarray}

So in this simple model, the mean of the ratio $R$ is precisely the diffuse flux divided by the total flux. The variance in $R$ is generally Poisson, scaling as $1/$(total photon counts) provided the number of counts per pixel from diffuse emission is $\lesssim 1$, but is reduced when the diffuse flux is much smaller than the total flux. (It is possible for the diffuse flux to be much less than the total flux even when most cells do not contain point sources, but not the converse, unless $\lambda$ is rather large, in which case the sample size or the cell size should be reduced.) The variance grows rapidly as $\lambda$ becomes greater than 1: if the cell size (corresponding to angular resolution) is too large relative to the amount of diffuse emission, $R$ is not a good measure of the fraction of diffuse flux (however, see \S \ref{sec:extension}).

If there is a significant contribution from weak point sources that may
only add a single isolated event to an empty cell (or alternatively,
if the cells are sufficiently small that this is common even for
stronger point sources), then the mean of $R$ is no longer given simply
by the fraction of diffuse flux. For example, suppose $S$ cells gain exactly one event when point sources are added, and $T$ cells in total are affected by the addition of point sources (so $T \ge S$): then the addition of point sources sends $p_0 \rightarrow (1 - T/N) p_0$ and $p_1 \rightarrow (1 - T/N) p_1 + (S/N) p_0$. Then we obtain,
\begin{equation} \langle R \rangle = \frac{N}{N \lambda + \Gamma} \left(\lambda + \frac{S}{N - T} \right), \end{equation}
\begin{equation} \sigma^2(R) = \left(\frac{1}{N \lambda + \Gamma} \right) \frac{e^\lambda \left( 1 + \lambda + \frac{S}{N-T} \right) \left(1 + \frac{S}{\lambda (N - T)} \right)}{\left(1 + \Gamma / N \lambda \right) \left(1 - T/N \right)}. \end{equation}
If the number of cells containing single 1-photon point sources is small compared to the number of cells unaffected by point sources ($S \ll N-T$), and also the number of photons from these weak point sources is small compared to the number of diffuse events from cells unaffected by point sources ($S \ll \lambda (N - T)$), we recover the previous result. If the second condition fails to hold (as occurs in the limit of low diffuse emission, independent of the point source luminosity function), then $\langle R \rangle$ asymptotes to (flux from isolated 1-photon point sources)/(total flux) $\times 1/(1 - T/N)$, as $\lambda \rightarrow 0$ with the total flux held constant. If this limiting value is $\gsim 1$, then $R$ has no discriminatory power, and the situation is not improved by higher statistics: this is simply the statement that there is no difference between diffuse emission and a very large number of very faint uniformly distributed point sources.

\subsection{Extension to the case of large $\lambda$}
\label{sec:extension}

One region of parameter space in which this test breaks down is where
$\lambda \ge 1$. However, a simple generalization of the statistic can
be useful in this case. Suppose we make a histogram of the number of
nearest neighbours each event possesses, and measure the peak of the
histogram to be some number of neighbours $n_\mathrm{crit}$. Let us
redefine $n_I$ as the fraction of events with $n_\mathrm{crit}$ or fewer
neighbours, and $n_E$ as the fraction of points with $n_\mathrm{crit}$ or
fewer events within the test radius $r$.  Then for Monte Carlo
realizations of diffuse flux plus some randomly distributed point
sources, we can again measure the ratio $n_I/n_E$ and use it as a
measure of how correlated the photon events are (see \S \ref{sec:deppsf}).

\section{The point source luminosity function}

Clearly, the sensitivity of the test depends critically on the fraction
of point sources which contribute at most one count to the data, which
is determined by the point source luminosity function. In the limit
where the point source flux is dominated by (a very large number of)
uniformly distributed sources which each produce an average number of
counts $\ll 1$, unresolved point sources are practically
indistinguishable from diffuse emission.

In the Monte Carlo tests which follow, we treat the luminosity function as some unbroken power law $dN / dS = S^{-\alpha}$ between integrated flux limits $S_\mathrm{min}$ and $S_\mathrm{max}$. $S_\mathrm{max}$ is bounded above by the faintest point sources which can be resolved and masked out. $S_\mathrm{min}$, on the other hand, is not determined by known properties of the experiment, and limits derived with a particular value of $S_\mathrm{min}$ should be interpreted as placing a limit on the contribution from point sources with average luminosity above $S_\mathrm{min}$. 

The effect of $S_\mathrm{min}$ on the behaviour of $R$ depends on the value of the spectral index $\alpha$: if $\alpha > 2$, then most of the flux originates from the faintest point sources, and the total flux diverges at low luminosity. In this case, the power law generally breaks to a much shallower slope at some low luminosity, meaning that most of the flux originates from point sources with luminosities close to the break: we can approximate this behaviour simply by cutting off the luminosity function at $S_\mathrm{min}$. If $\alpha < 2$ then most of the flux is concentrated in the brighter point sources and changing $S_\mathrm{min}$ has little effect on $R$.

To estimate the relevant range of $\alpha$ we examine studies of known populations of point sources. The gamma-ray luminosity functions of AGN contributing to the extragalactic gamma ray background have been studied using the first three months of data from the \Fermi LAT \citep{Abdo:2009wu}. The luminosity function for BL Lac objects was found to be well described by a single power law with $\alpha = 2.17 \pm 0.05$. The luminosity function for flat spectrum radio quasars (FSRQs) was well described by a power law with $\alpha = 2.58 \pm 0.19$ at high redshifts ($z \ge 1$), indicating that at high redshifts the \Fermi LAT is sampling the bright (steep) end of the luminosity function, but at $z \le 1$ the best-fit value of the slope was $\alpha = 1.56 \pm 0.10$. The X-ray luminosity functions of the same classes of objects have been studied by \cite{Padovani:2007qb}, with BL Lac objects measured to have $\alpha = 2.12 \pm 0.16$, and $\alpha = 1.6-1.9$ for FSRQs at $z \le 1$.

The X-ray luminosity function of high-mass X-ray binaries (HMXBs) was found by \cite{Grimm:2002ta} to have slope $\alpha = 1.61 \pm 0.12$. Studies of low-mass X-ray binaries (LMXBs) in Centaurus A \citep{Voss:2005aq, 2009ApJ...701..471V} yield a slope of $\alpha = 1.8-2.0$ at high luminosity, flattening to $\alpha \sim 1.2$ at low luminosity, in agreement with earlier studies of LMXBs \citep{Kim:2003tr, Gilfanov:2003th}.

In the examples in the following section, therefore, we take our two benchmark models to have $\alpha = 1.8, 2.2$, and also demonstrate the effect of varying $\alpha$ between $1.5$ and $3.0$. Smaller values for $\alpha$ improve the ability of our statistic to distinguish point sources from diffuse emission, simply because more of the flux originates from bright sources.

\section{Monte Carlo examples}
\label{sec:mc}

To examine the usefulness of the ratio $R$ as a measure of the point source flux, we employ a Monte Carlo approach. We consider a given angular ``signal region'', and a smaller ``signal window'' within that region (the purpose of this distinction is to eliminate edge effects). Within the signal region we generate a uniform random distribution of point source locations, for a certain number of sources (which we vary, as a proxy for varying the total flux from point sources). The expected number of counts for each point source is drawn from a power law distribution with spectral index $\alpha$, with cutoffs at a minimum expected number of counts $S_\mathrm{min}$, and a maximum expected number of counts $S_\mathrm{max}$. For our benchmark models we take $(1)$ a ``pessimistic'' set of parameters $\alpha = 2.2$, $S_\mathrm{max} = 10$, $S_\mathrm{min} = 0.1$, and $(2)$  an ``optimistic'' set of parameters $\alpha = 1.8$, $S_\mathrm{max} = 100$, $S_\mathrm{min} = 1$ (see Table \ref{tab:benchmarks}). Once the expected counts from each source have been obtained, the number of counts actually observed from each source is determined by a Poisson draw; their angular distribution is determined by the detector PSF, which for simplicity we model as a Gaussian with standard deviation $\sigma$.

We compute the total number of counts lying within the signal window and originating from the point sources, and subtract this quantity from the desired total number of counts in the signal window. The result approximates the number of diffuse events we wish to generate in the signal window. We then generate a uniform random distribution of diffuse events in the signal region, with the total number of events given by the desired diffuse counts in the signal window rescaled to the greater area of the signal region. The total photon distribution in the signal window (diffuse + point sources) then has approximately the correct number of events. We can also incorporate non-uniformities in the diffuse flux at this point.

Having produced our test data, we need to compute the ratio $R$. For each event within the signal window, we use the publicly available IDL routine \texttt{spherematch} \footnote{The IDL routines used in this analysis are available as part of v5\_4\_8 of the IDLUTILS product at \texttt{http://sdss3data.lbl.gov/software/idlutils}} to find its neighbours in the set of events in the larger signal region. This eliminates edge effects, i.e. a spuriously high number of isolated events at the edges of the signal window. We repeat the process for a random distribution of points within the signal window, to compute $n_E$, and plot the resulting $R$ against the fraction of flux in the signal window originating from point sources. 

For these examples, we take the default signal window to be the region
$|l| < 15$, $|b| < 15$, the PSF to be $0.2^\circ$, and the total number
of counts to be 3000, corresponding to a mean $\lambda = 0.4$ events per
PSF circle. Note that the statistic is insensitive to the shape of the
signal window, and if we rescale the PSF by some factor $a$ and the area
of the signal window by $a^2$, and hold the number of events constant,
then this is just equivalent to a unit redefinition and does not change
the results. 

When we refer to the ``PSF'', we mean the standard error $1 \sigma$.
Optical astronomers often use the full width at half max (FWHM), which
for a Gaussian equals $2.355\sigma$ and contains 50\% of the flux.
Gamma-ray astronomers often use the radius of 68\% or 95\% containment
($1.51\sigma$ or $2.45\sigma$, respectively).  The $R$ statistic is not
greatly sensitive to the exact choice of $r$.  We will demonstrate the
effect of varying the PSF while holding the other parameters constant
(thus changing $\lambda$, the mean number of events per PSF circle), and
of increasing the number of counts while holding $\lambda$ constant, by
increasing the signal window. We will also generally assume that the
test radius $r$ is equal to $1\sigma$, but show the effect of using a
different test radius, for both benchmark luminosity functions. 

\begin{figure*}
\includegraphics[width=0.45\textwidth]{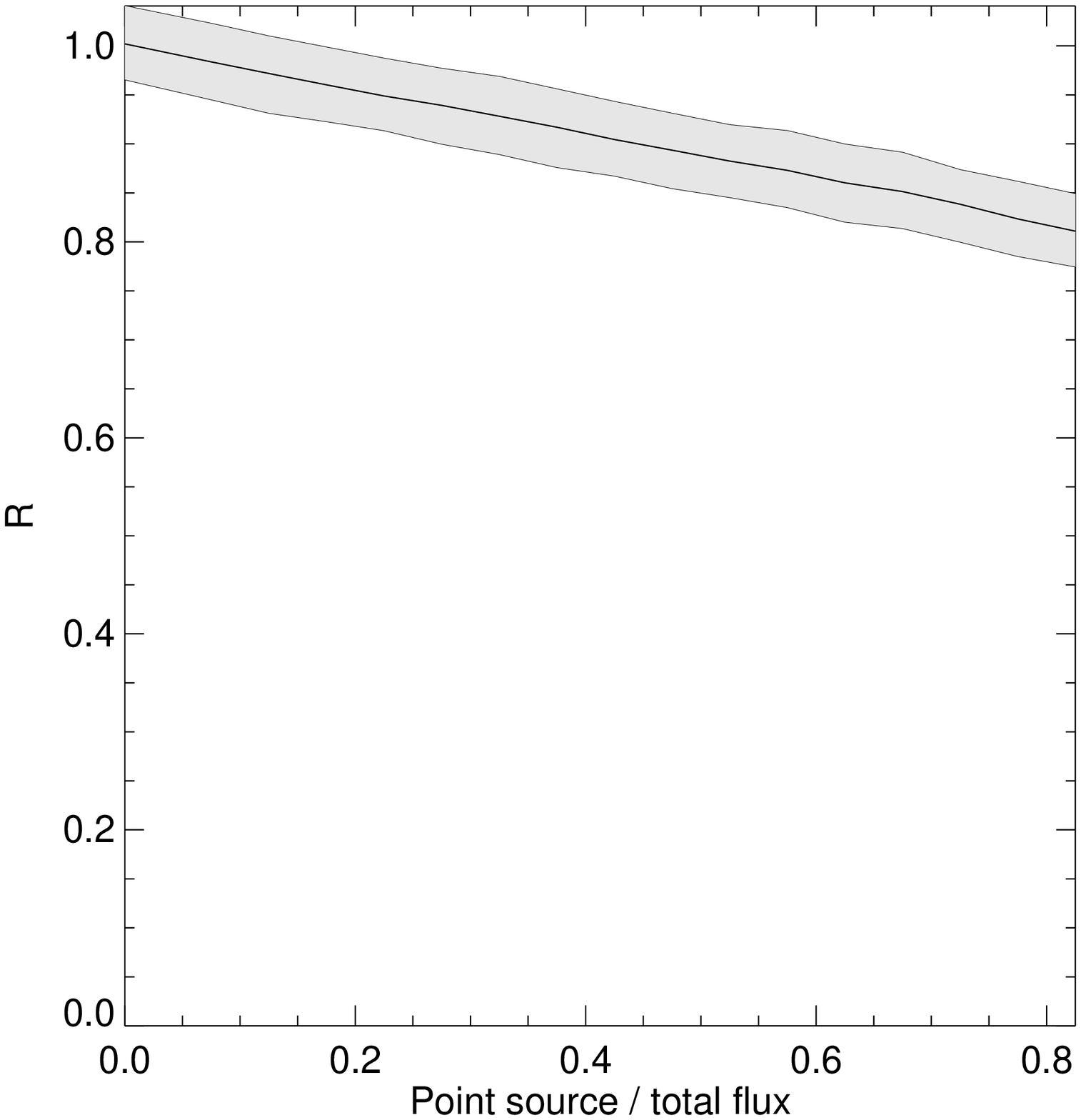}
\includegraphics[width=0.45\textwidth]{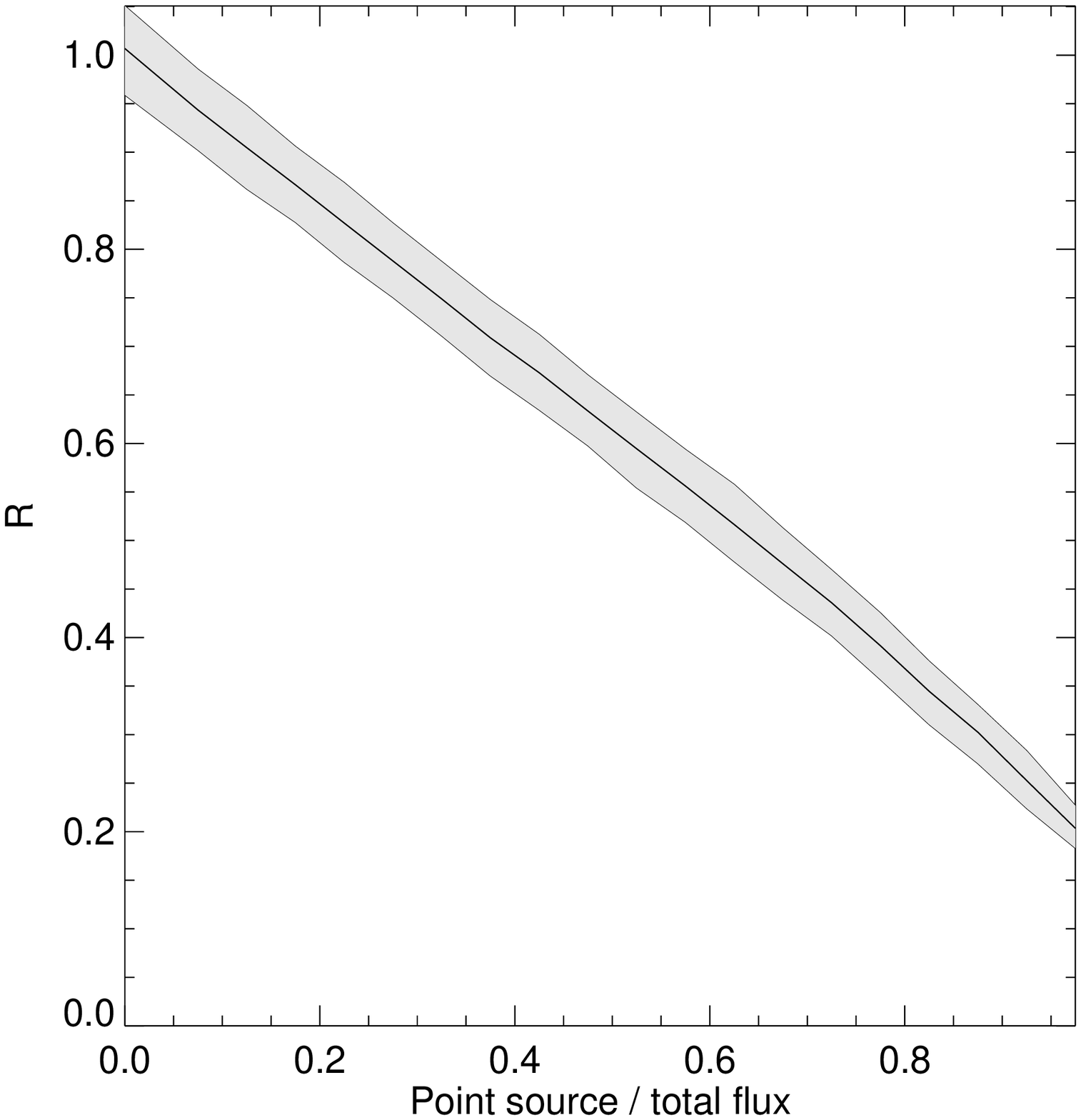}
\caption{\label{fig:benchmarks} 
The isotropy ratio $R$, in Monte Carlo simulated data, in (\emph{left})
benchmark model 1 and (\emph{right}) benchmark model 2.  See Table
\ref{tab:benchmarks} for definitions of the benchmarks.  Lines are 5\%
and 95\% confidence bounds.}
\end{figure*}

In the plots that follow, we will describe the ``sensitivity'' of this test by two representative measures:
\begin{enumerate}
\item The maximum value of the point source flux fraction consistent with $R=1$, within the $95 \%$ confidence limits.
\item Bounds on the point source flux fraction when the true point source contribution is half the signal, obtained by averaging the limits which would be obtained from an R-measurement over the histogram of $R$-values produced in this scenario.
\end{enumerate}
The statistic is most powerful where the bound in (i) is small and the limits in (ii) are close together. If either the variance of $R$ (for fixed point source flux) becomes large, or the mean of $R$ is slowly varying with respect to the point source flux fraction, then both of these measures will blow up. We also overplot the sensitivity estimates obtained from the analytic approximation for $R$ in the grid model described in \S \ref{sec:analytic}, using the relations in Appendix \ref{app:lfanalytic}, for comparison to the results of the MC runs.

\begin{table}
\caption{Benchmark parameters for luminosity functions $dN/dS\sim
  S^{-\alpha}$ with $S_\mathrm{min} < S < S_\mathrm{max}$.}
\label{tab:benchmarks}
\begin{tabular}{@{}lccc}
\hline
               & $\alpha$ & $S_\mathrm{max}$  & $S_\mathrm{min}$ \\
\hline
Benchmark 1    & 2.2        & 10              & 0.1              \\
Benchmark 2    & 1.8        & 100             & 1                \\
\hline
\end{tabular}
\end{table}

\section{Results}
\label{sec:results}

We find a strong linear relationship between $R$ and the fraction of the flux due to diffuse emission, as shown in Fig. \ref{fig:benchmarks} for the two benchmark sets of parameters. As expected, the ``optimistic'' benchmark parameters render $R$ more sensitive to the fraction of diffuse emission.

\subsection{Dependence on the point source luminosity function and the effect of improved statistics}

\begin{figure*}
\includegraphics[width=0.24\textwidth]{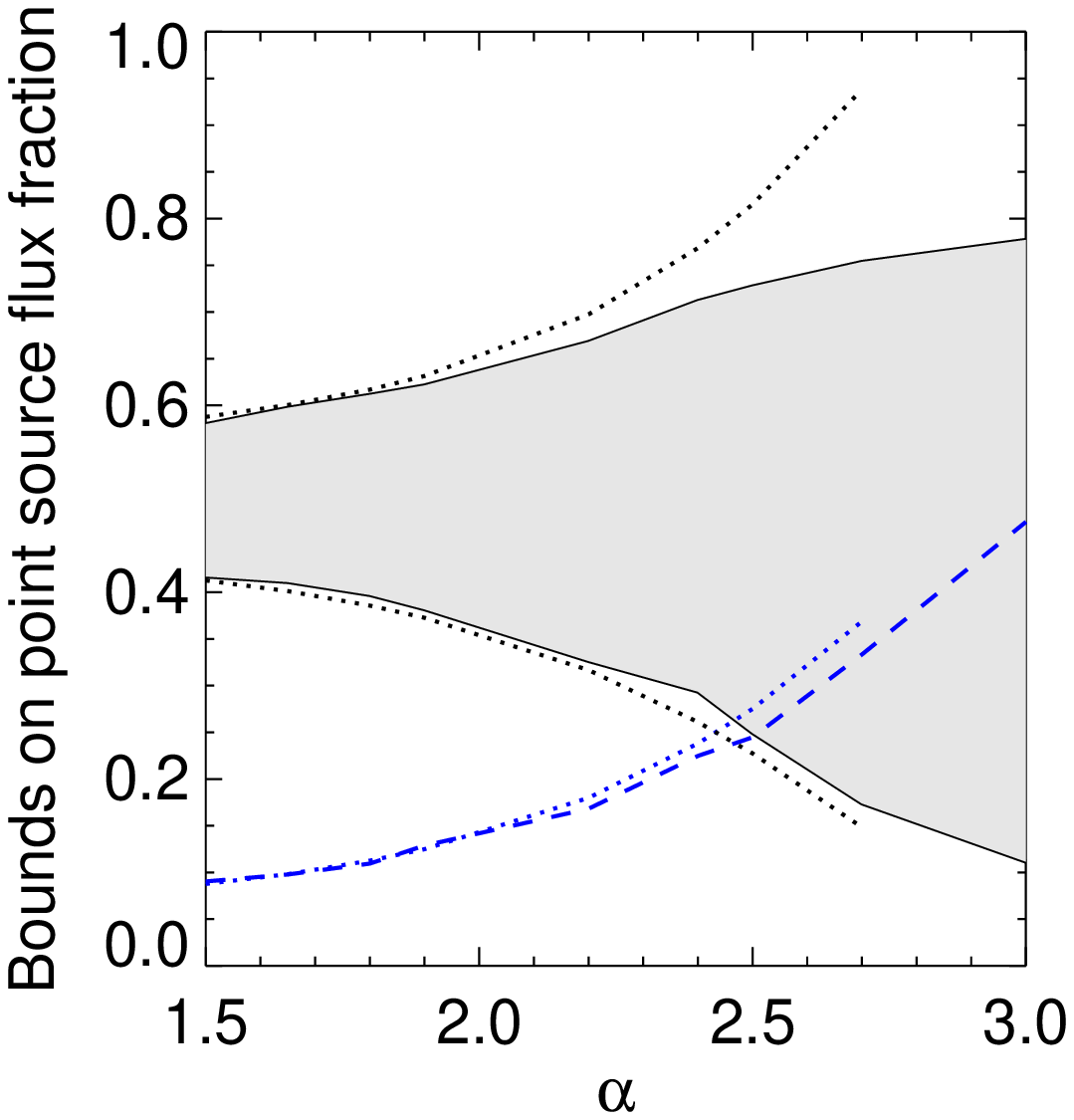}
\includegraphics[width=0.24\textwidth]{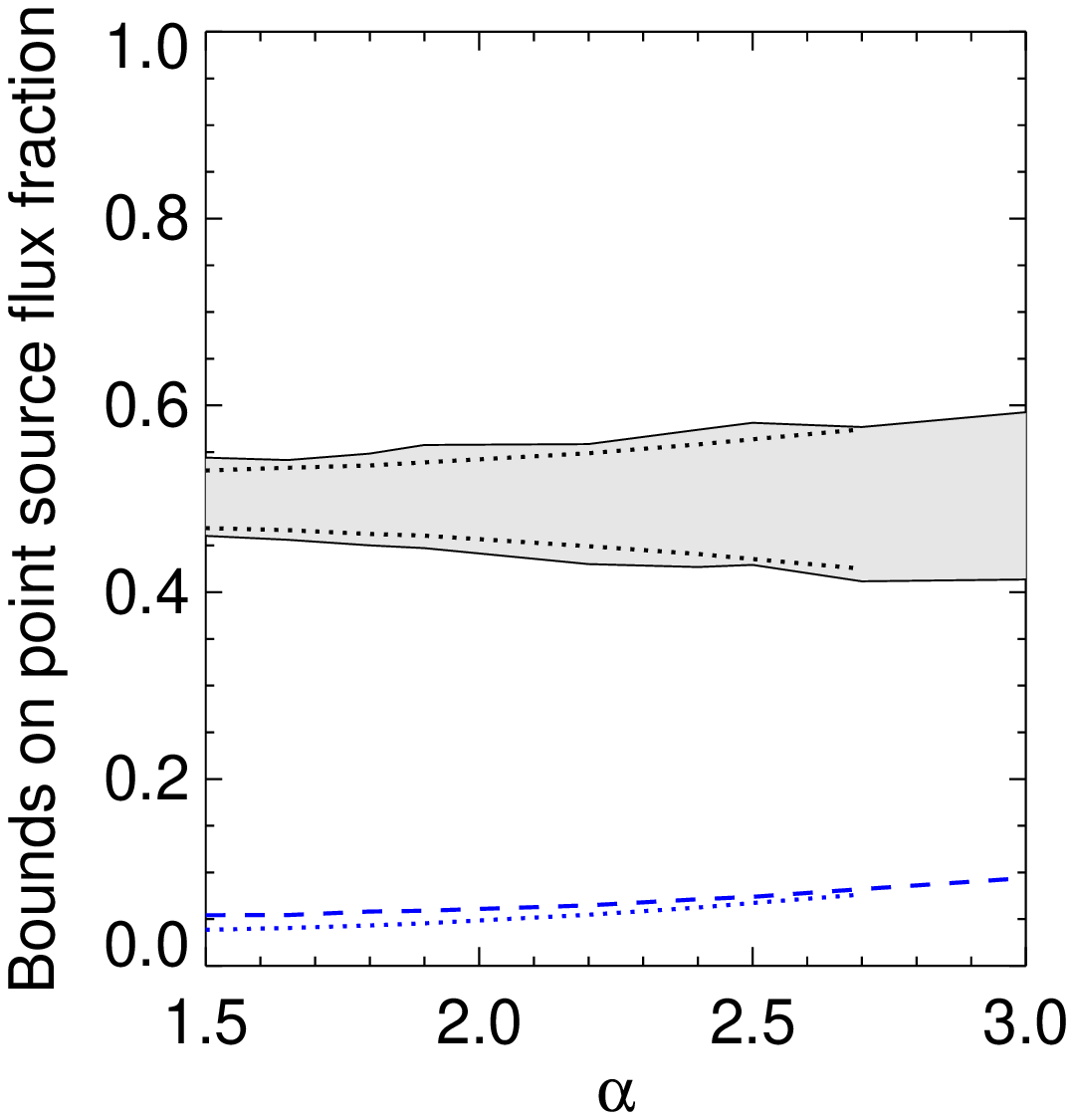}
\includegraphics[width=0.23\textwidth]{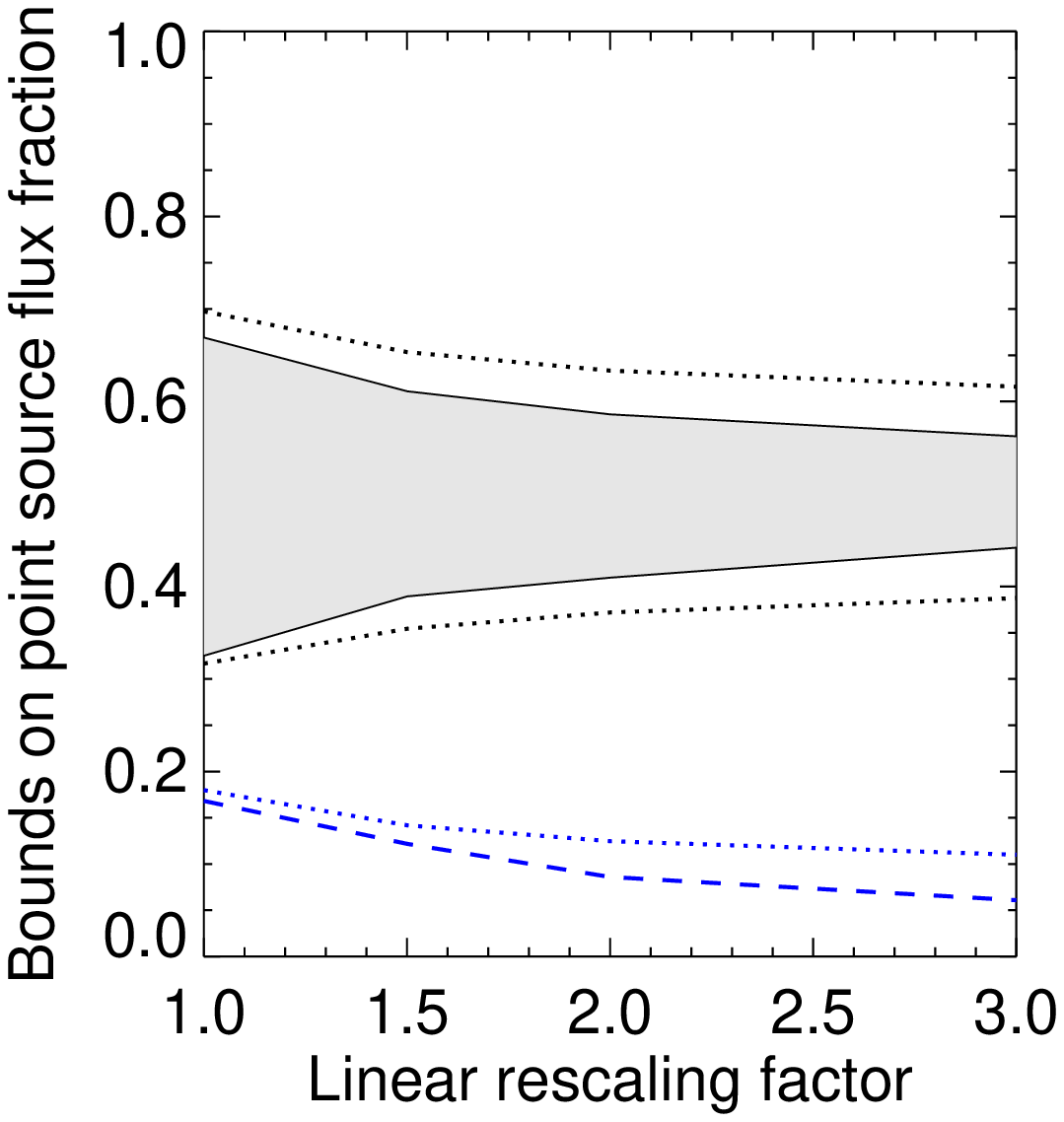}
\includegraphics[width=0.23\textwidth]{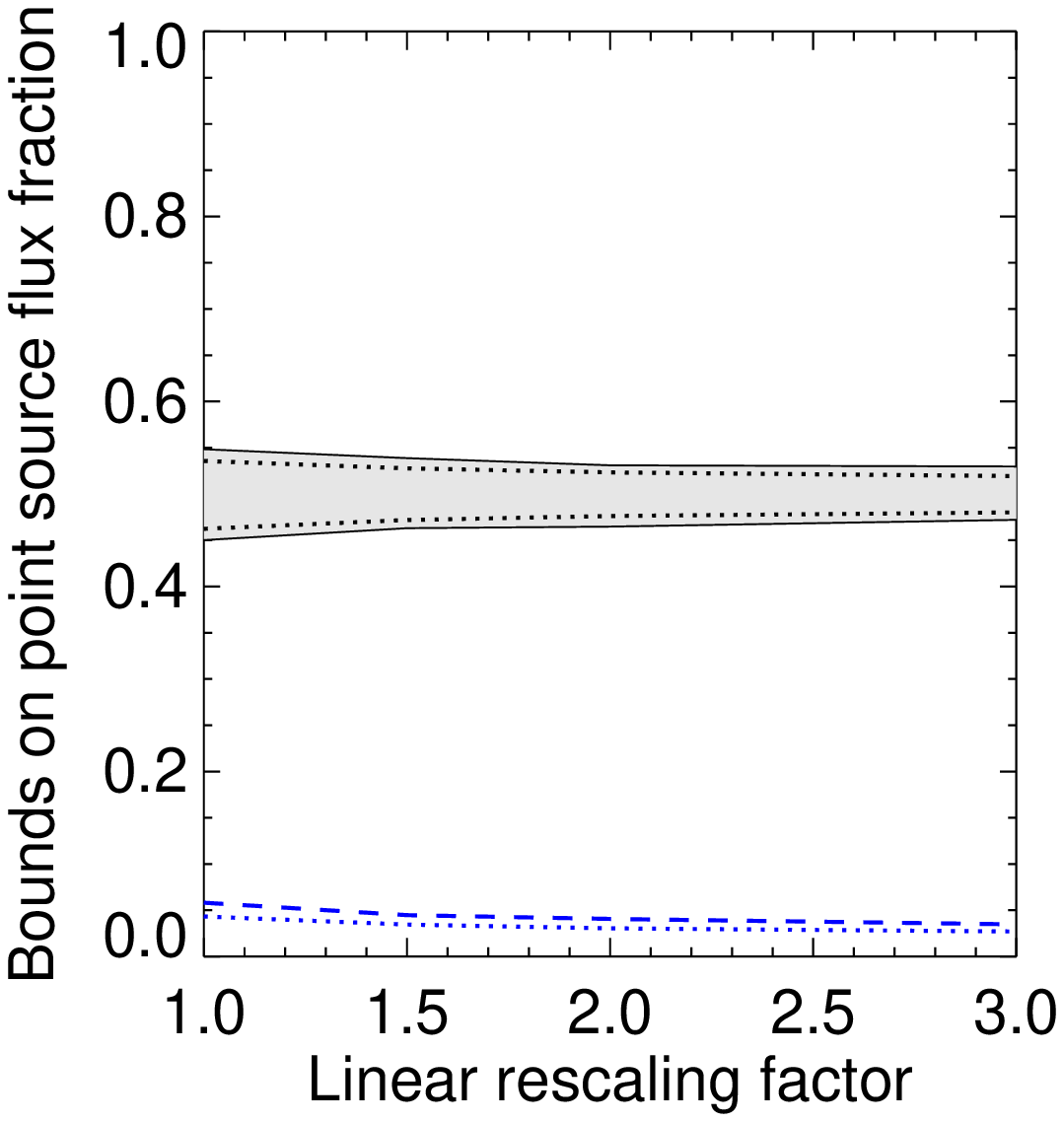}
\caption{\label{fig:sdependence} 
The sensitivity of $R$ as a function of the spectral index of the point source luminosity function, $\alpha$ (\emph{far left:} benchmark model 1, \emph{center left:} benchmark model 2), and as a function of the signal area while the density of counts is held constant (\emph{center right:} benchmark model 1, \emph{far right:} benchmark model 2). As described in the text, solid lines (bounding the shaded area) indicate the average 90 $\%$ confidence bounds on the point source flux fraction from this test when the true fraction is 0.5; the dashed line indicates the $95 \%$ confidence upper limit on the point source flux fraction where $R=1$. Dotted lines indicate the analogous results for the analytic approximate calculation.}
\end{figure*}

Fig. \ref{fig:sdependence} shows the effect of varying the spectral index of the point source luminosity function, for the two benchmark choices of $S_\mathrm{min}$, $S_\mathrm{max}$. As expected, smaller values of $\alpha$ yield better performance for $R$ as a precise estimator of the diffuse flux. For the pessimistic benchmark model, when $\alpha \sim 3.0$ the method has lost most of its discriminatory power: however, for more well-motivated power laws $\alpha \lesssim 2.5$, useful limits can still be obtained. For the optimistic benchmark model, $R$ can be used to place strong limits on the diffuse emission even for very steep power laws $\alpha \sim 3.0$, largely because the assumed break in the power law at $S_\mathrm{min} = 1$ avoids the scenario of emission dominated by many very faint point sources ($S \ll 1$).

As a check on our understanding of the method, we can increase the area of the signal window and also the number of counts, holding the density of events constant. This change would be expected to have no effect other than reducing the variance in $R$, since $\sigma(R)$ is expected to scale as $1/\sqrt{\text{number of counts}}$. Fig. \ref{fig:sdependence} demonstrates the improvement in sensitivity.

\subsection{Dependence on the PSF and test radius $r$}
\label{sec:deppsf}
\begin{figure*}
\includegraphics[width=0.24\textwidth]{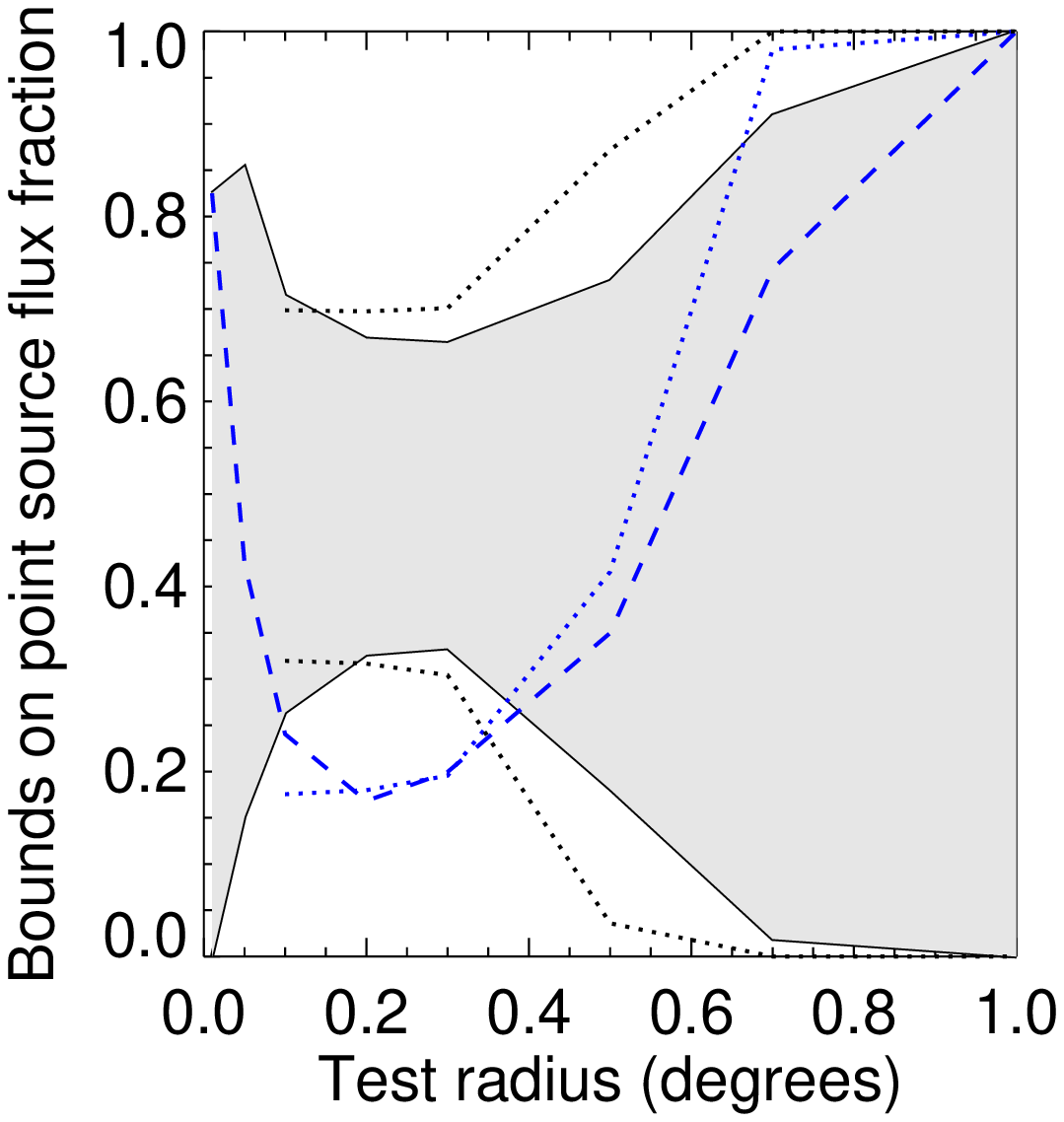}
\includegraphics[width=0.24\textwidth]{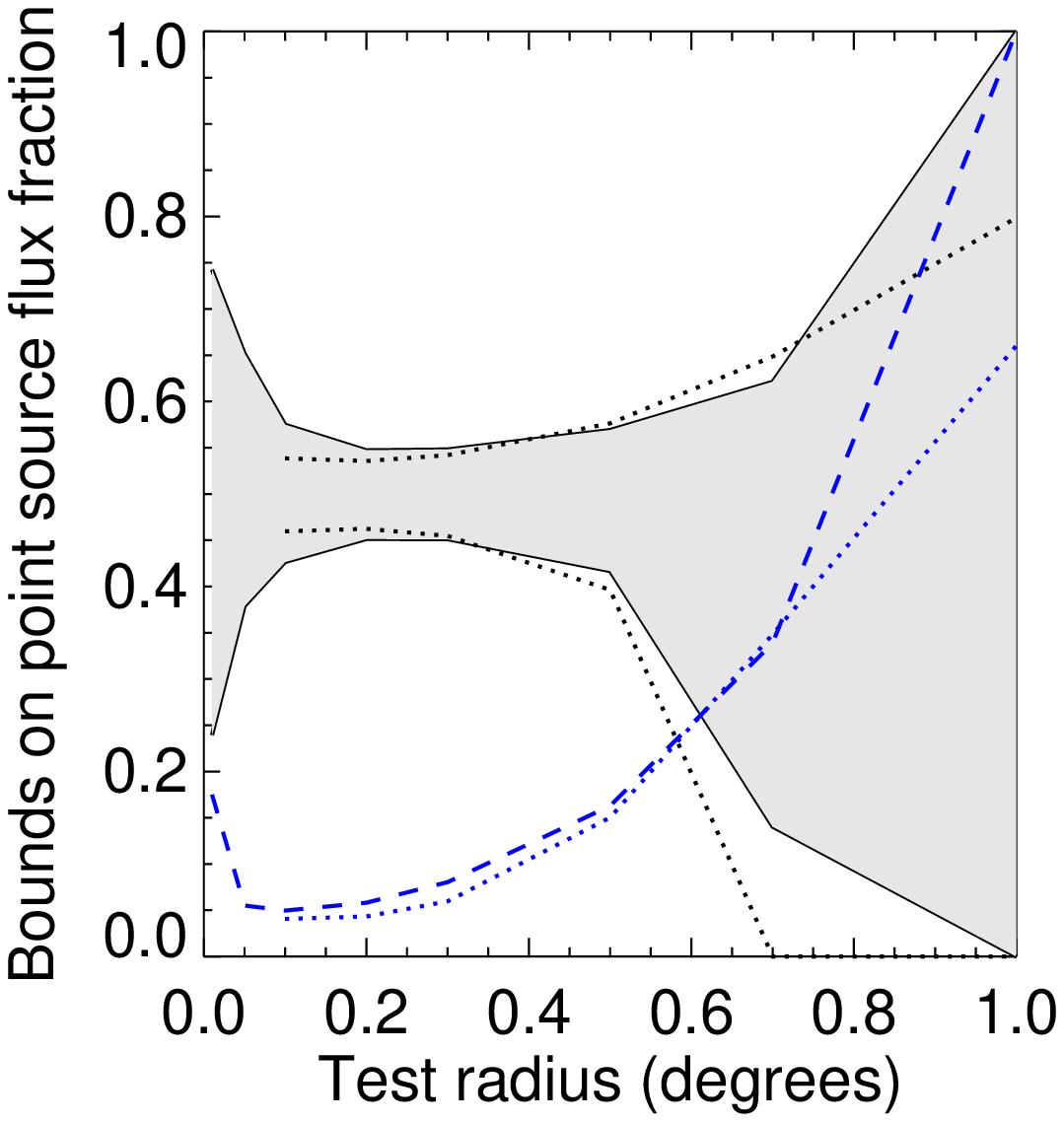} 
\includegraphics[width=0.24\textwidth]{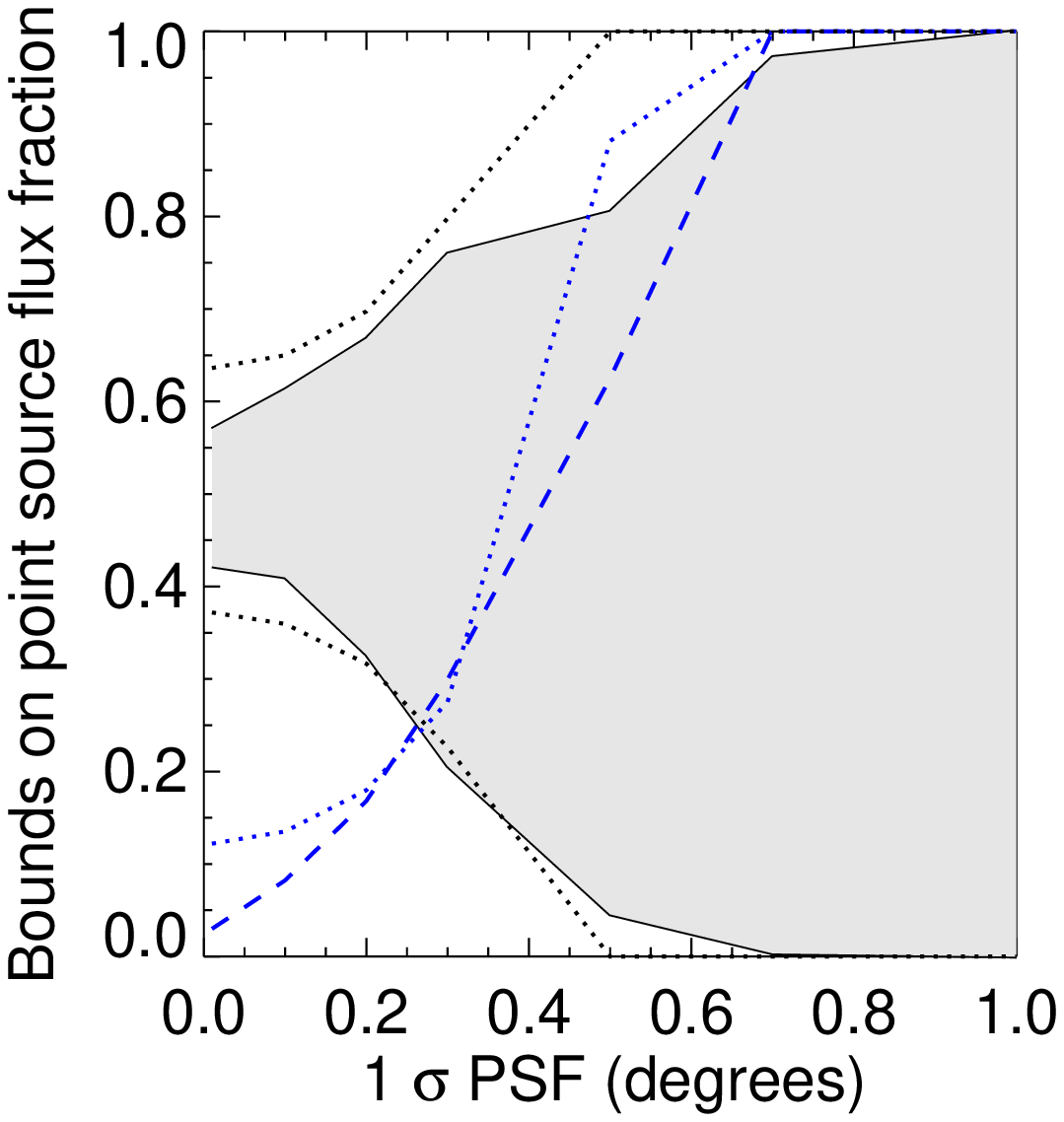}
\includegraphics[width=0.24\textwidth]{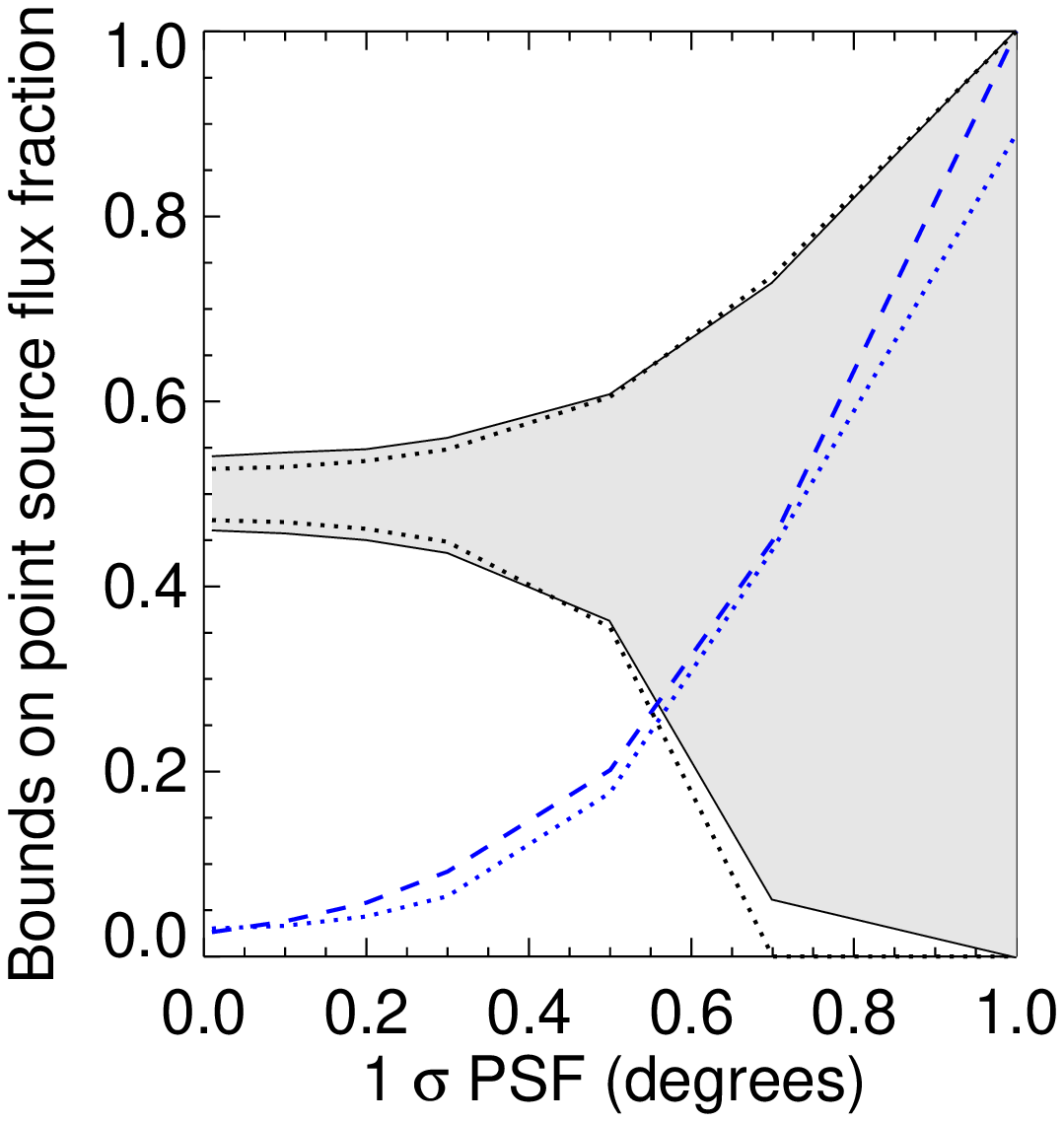} \\
\includegraphics[width=0.24\textwidth]{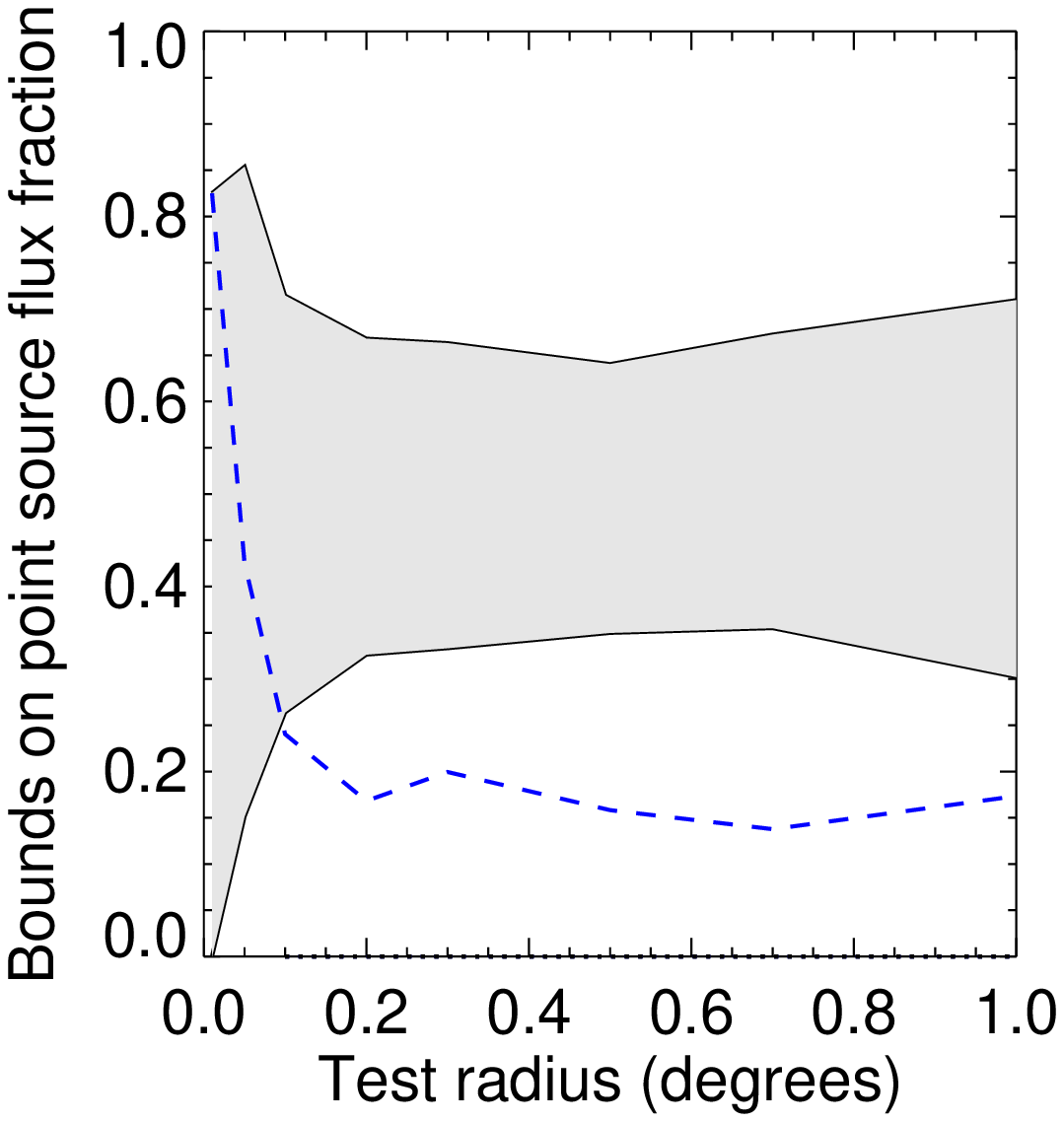}
\includegraphics[width=0.24\textwidth]{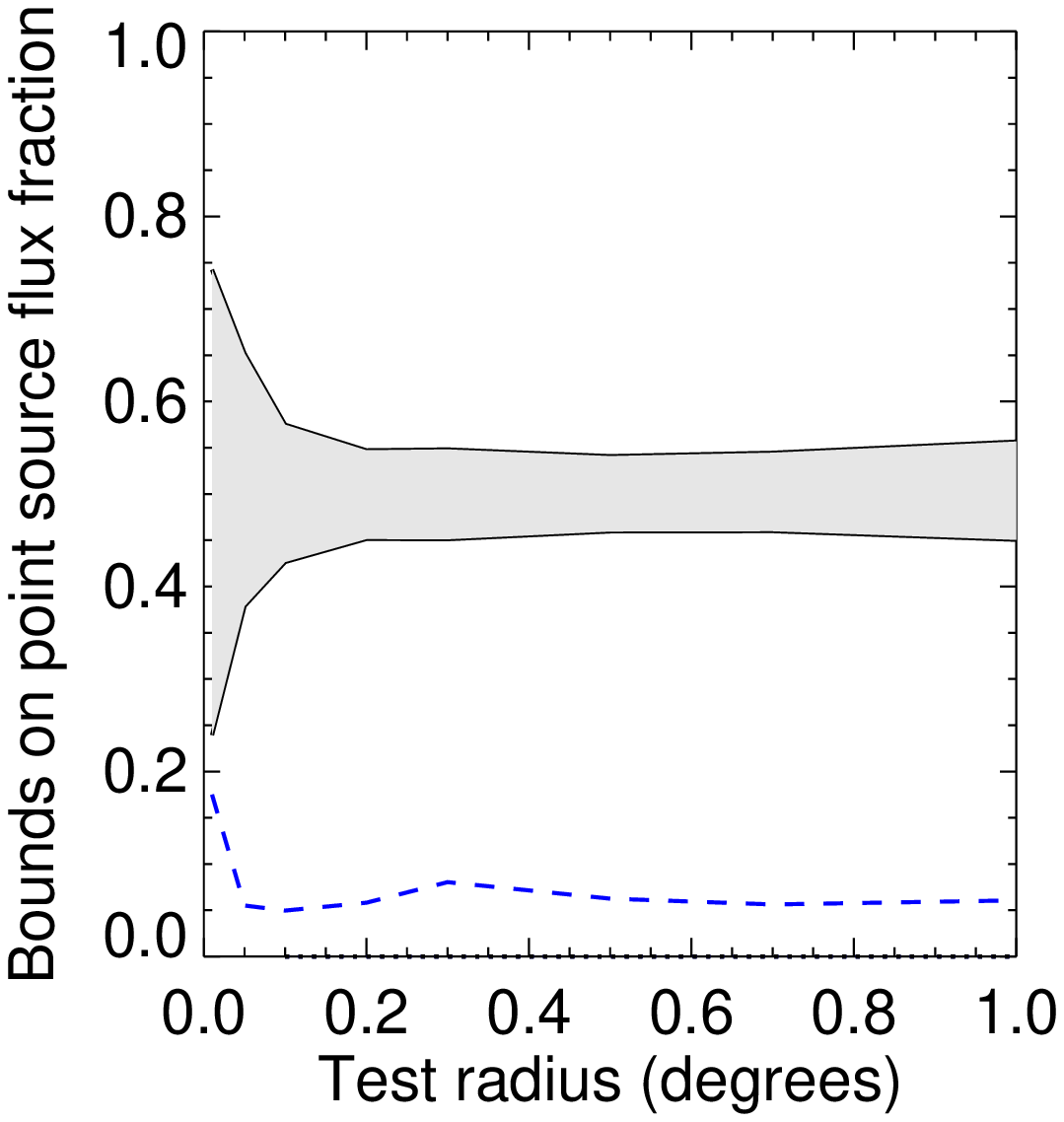}
\includegraphics[width=0.24\textwidth]{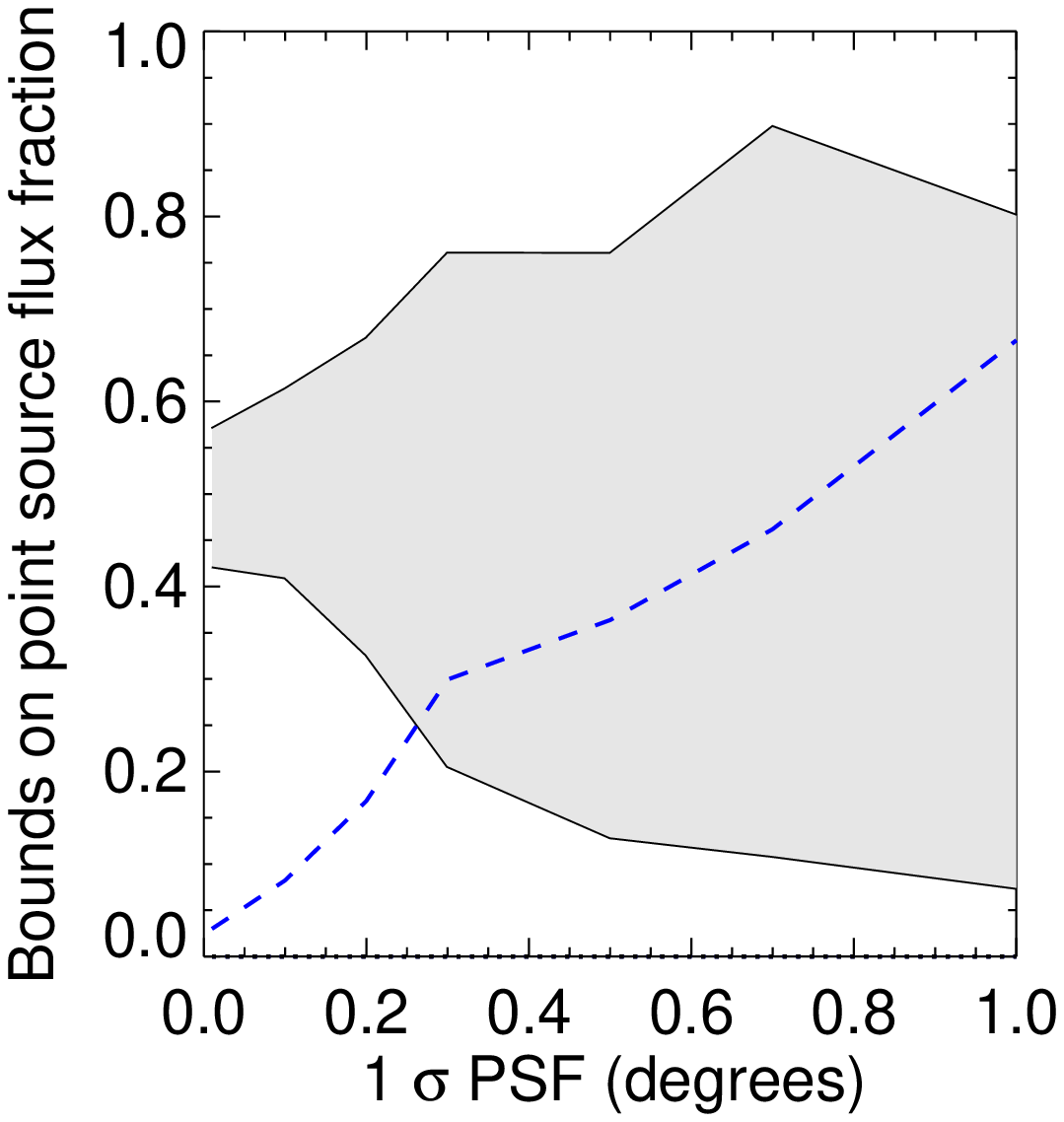}
\includegraphics[width=0.24\textwidth]{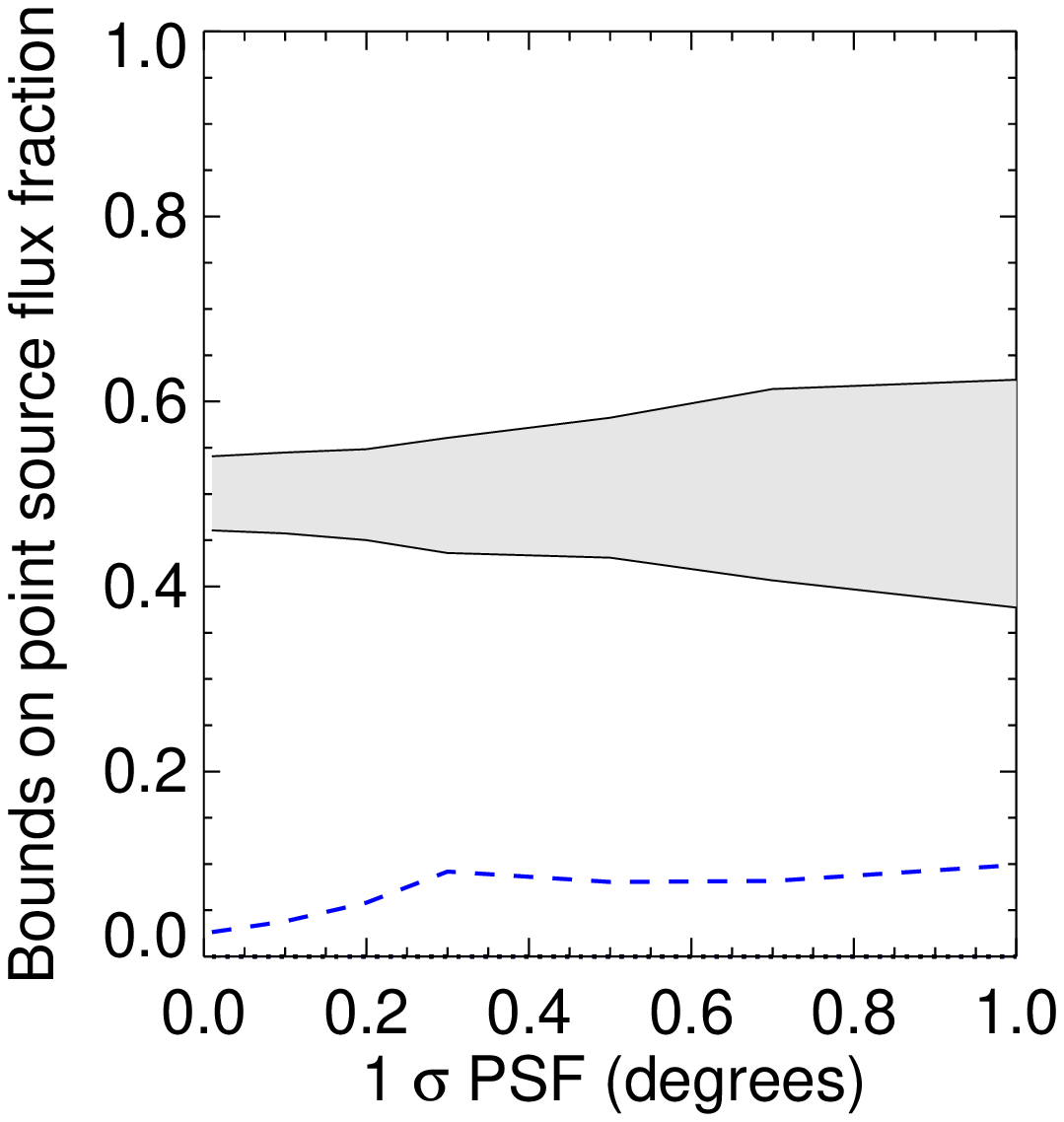}
\caption{\label{fig:varyingr} 
The sensitivity of $R$ as a function of the test radius $r$, holding the PSF $1 \sigma$ constant at $0.2^\circ$ (\emph{far left:} benchmark model 1, \emph{center left:} benchmark model 2), and as a function of the PSF size, fixing the test radius $r$ to be equal to $1 \sigma$ for the PSF (\emph{center right:} benchmark model 1, \emph{far right:} benchmark model 2). The top row uses the standard form of the statistic whereas the bottom row uses the modified form (\S \ref{sec:extension}) to prevent the catastrophic failure at large $r$. As described in the text, solid lines (bounding the shaded area) indicate the average 90 $\%$ confidence bounds on the point source flux fraction from this test when the true fraction is 0.5; the dashed line indicates the $95 \%$ confidence upper limit on the point source flux fraction where $R=1$. In the top row, dotted lines indicate the analogous results for the analytic approximate calculation; note that the truncation approximation made to derive the analytic result breaks down for $r \ll 1 \sigma$ PSF.}
\end{figure*}

The choice of the test radius $r$ significantly impacts the quality of the results. If $r$ is increased sufficiently that the expected number of neighbours for each event is $\gg 1$, then the number of isolated events and empty circles both become small in the diffuse limit, and the Poisson fluctuations in $R$ become very large. In the limit where $r \ll$ PSF, even events in bright point sources may qualify as ``isolated'' and $R$ loses its power to discriminate between point sources and diffuse signal. In the large-$r$ case, a better result with much less noise can be obtained by the generalized method discussed in \S \ref{sec:extension}. These effects are displayed in Fig. \ref{fig:varyingr}. 

The range of $r$ in which $R$ provides a precise estimate of the point source flux is given approximately by,
\begin{equation} \text{PSF} \lsim r \lsim \sqrt{\frac{\text{area of window}}{\pi \times \text{number of events}}}. \label{eq:rlimit} \end{equation}
If this range is large, $r$ can be varied substantially without much adverse impact on the performance of $R$ as a measure of diffuse flux. For the parameters employed here, the permitted range of $r$ is quite narrow and the effects discussed above are pronounced. 

Where this allowed range vanishes, as the average number of counts per PSF circle exceeds 1, this method breaks down as discussed in \S \ref{sec:analytic}, and we should instead employ the generalization discussed in \S \ref{sec:extension}. Fig. \ref{fig:varyingr} shows the slight improvement in the sensitivity of $R$ in the case of a smaller PSF, the breakdown in the method at large PSF, and the performance of the generalized method.

\subsection{Inhomogeneity in the diffuse flux}

\begin{figure*}
\includegraphics[width=0.23\textwidth]{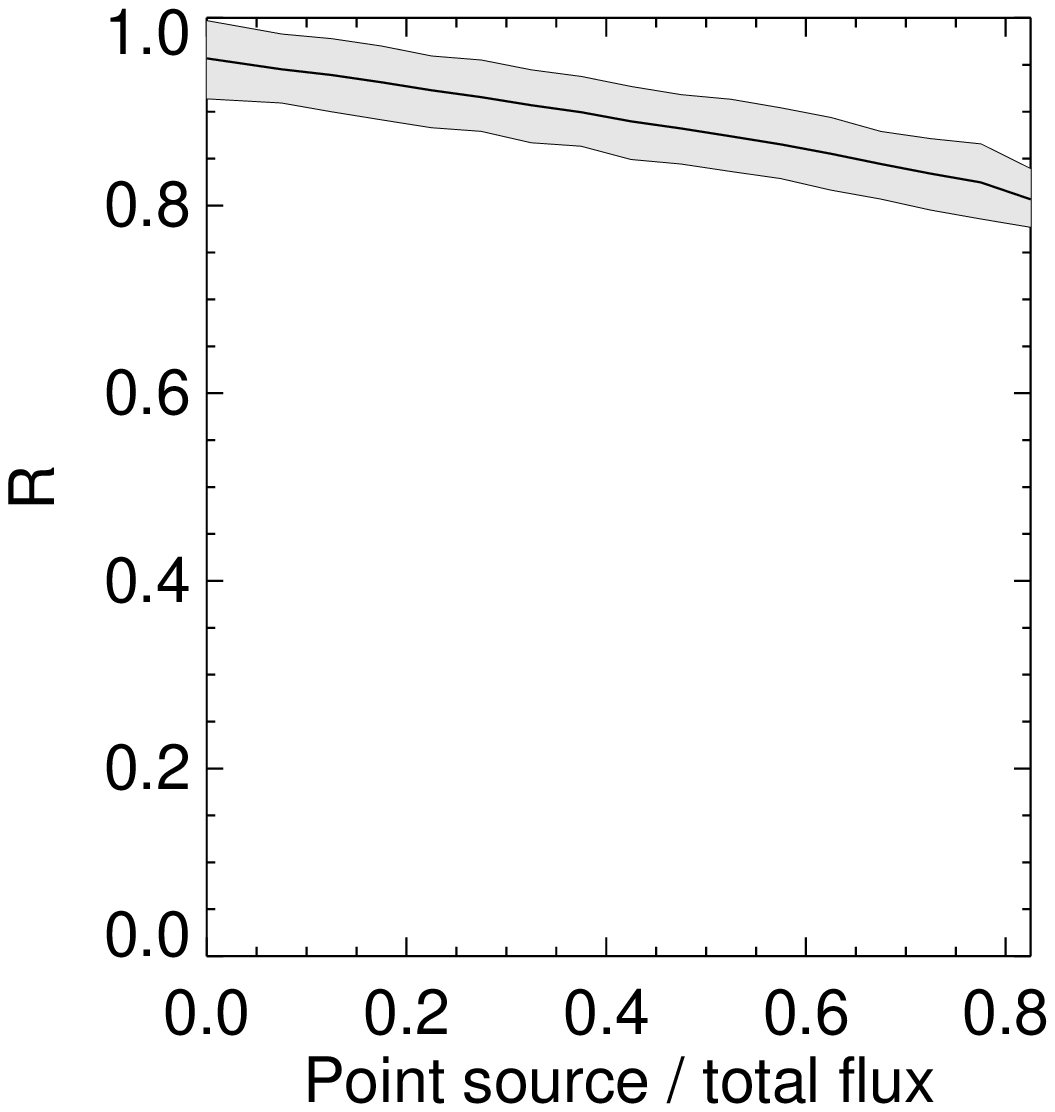}
\includegraphics[width=0.23\textwidth]{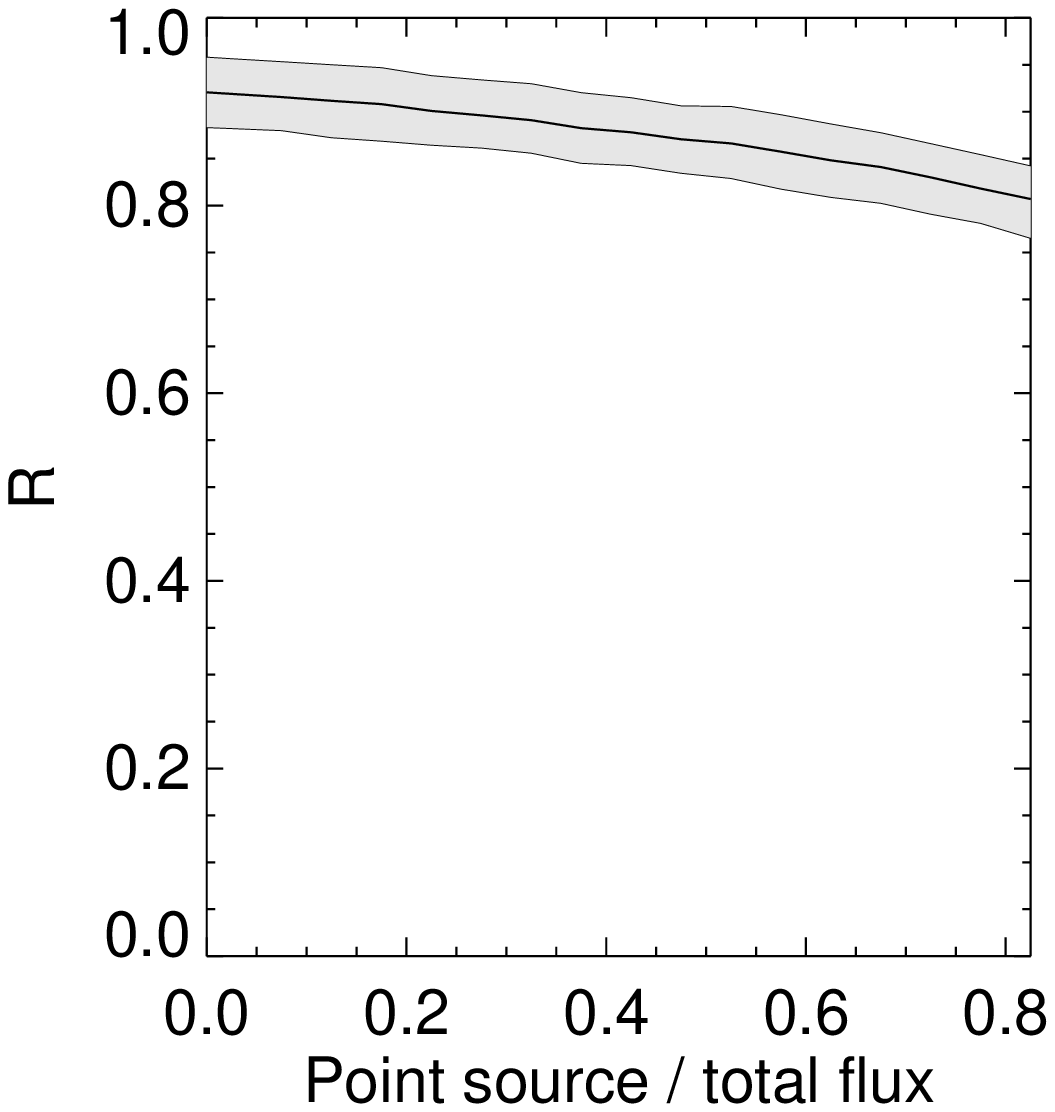}
\includegraphics[width=0.23\textwidth]{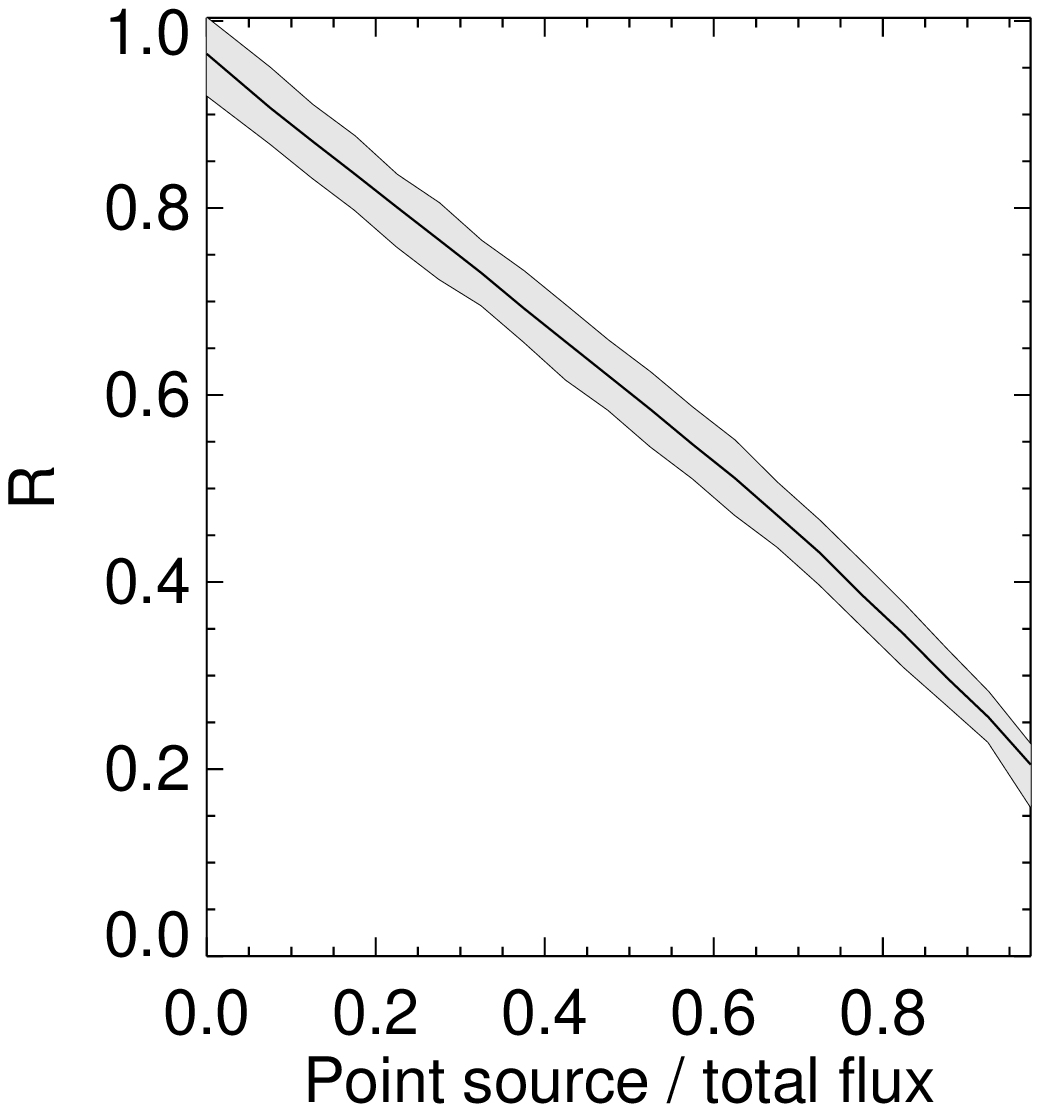}
\includegraphics[width=0.23\textwidth]{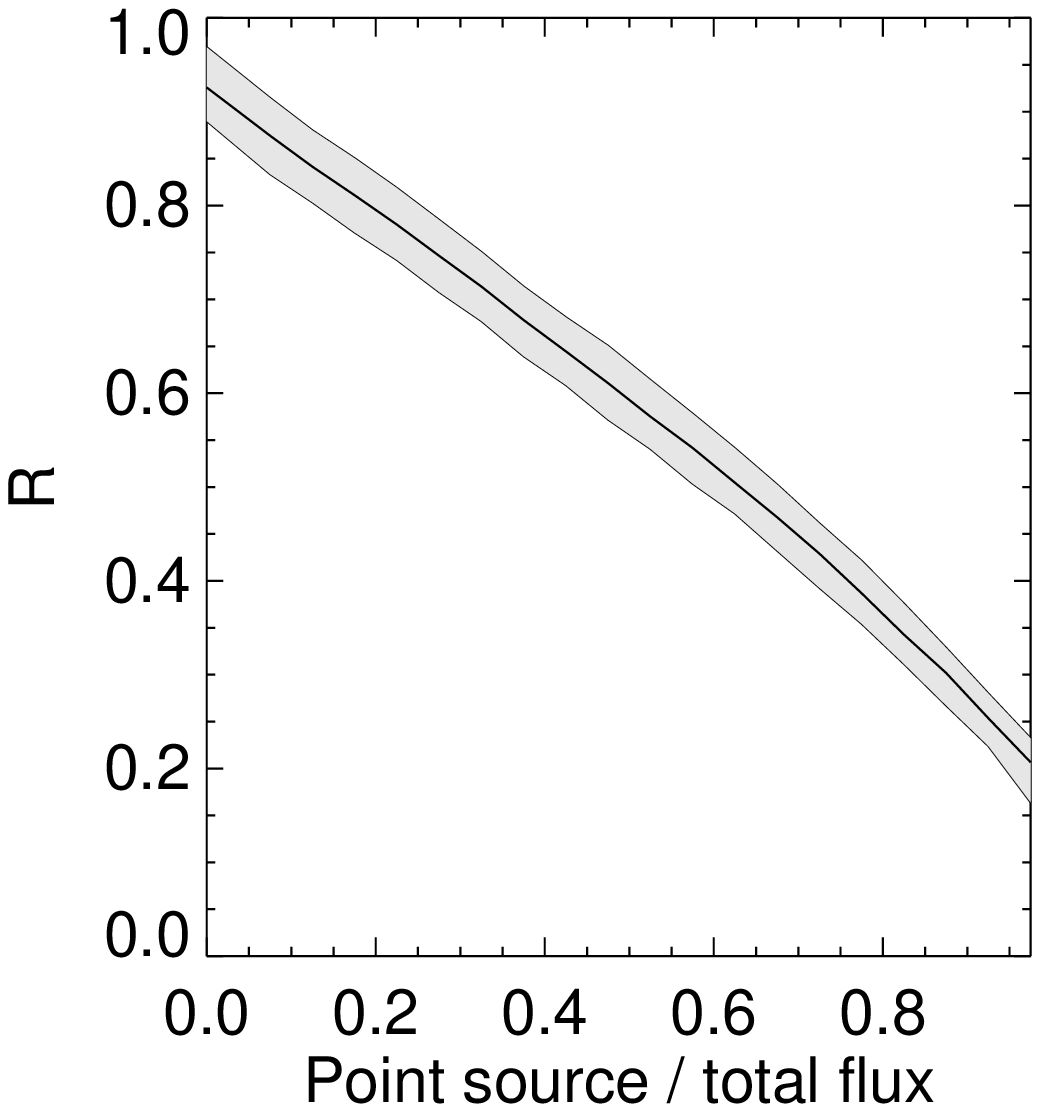}
\caption{\label{fig:varyingtilt} 
The effect of adding a $b$-dependent tilt to the diffuse photon flux, in Monte Carlo simulated data, for (\emph{far left}) benchmark model 1, tilt factor 5, (\emph{center left}) benchmark model 1, tilt factor 20, (\emph{center right}) benchmark model 2, tilt factor 5 and (\emph{far right}) benchmark model 2, tilt factor 20.}
\end{figure*}

The ``diffuse'' part of the sample may vary spatially, either because of
true spatial variation of the signal, or a non-uniform instrumental
sensitivity or exposure.  We show that $R$ is insensitive to such
variations by introducing a tilt in the diffuse photon distribution of
up to a factor of $20$ (that is, the density of events is 20 times lower
at one edge of the signal region than at the other) and observing that
it has little effect on $R$, as shown in Fig. \ref{fig:varyingtilt}. For large tilts, there is a noticeable downward shift in $R$ in the diffuse
limit.

\section{Comparison to the two-point function}

In essence, $R$ is a simple measure of the angular correlations between event positions, so it is reasonable to ask how it differs from the two-point correlation function. Consider the case of a single bright point source, compared to two (well separated) point sources with half the luminosity. The two-point correlation functions for these two situations are quite different, although the total flux from the point sources is the same, because the number of pairs in a given source scales as flux squared. The estimator $R$, on the other hand, to a first approximation does not probe correlations inside the point sources, and is dependent only on the fraction of diffuse flux, rather than the details of the sources.

This effect tends to reduce the variance of $R$, relative to the two-point function, as the flux from point sources increases and the luminosity function becomes shallower (i.e. brighter point sources, with (no. of counts)$^2 \gg$ no. of counts, contribute more of the signal); the $R$ statistic is also less sensitive to the luminosity function parameters than the two-point function. In the limit where the annulus width is large and almost all the non-isolated events have only a single neighbor, i.e. they form a single pair in the calculation of the two-point function (which is the case when the luminosity function is steep and the point source flux is dominated by faint sources, or when there are simply very few point sources), the two statistics capture essentially the same information and their performance is very similar.

\begin{figure*}
\includegraphics[width=0.3\textwidth]{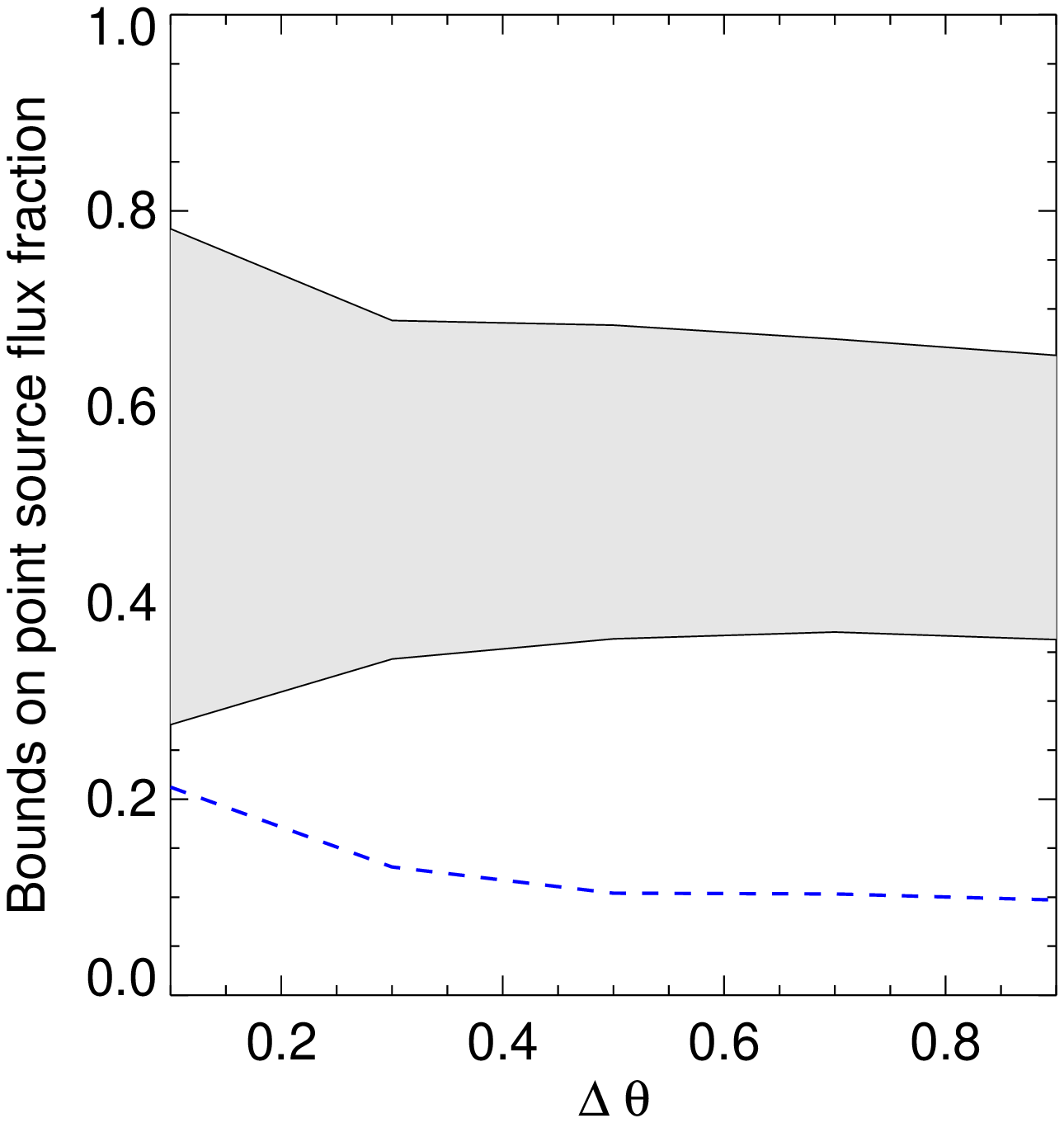}
\includegraphics[width=0.3\textwidth]{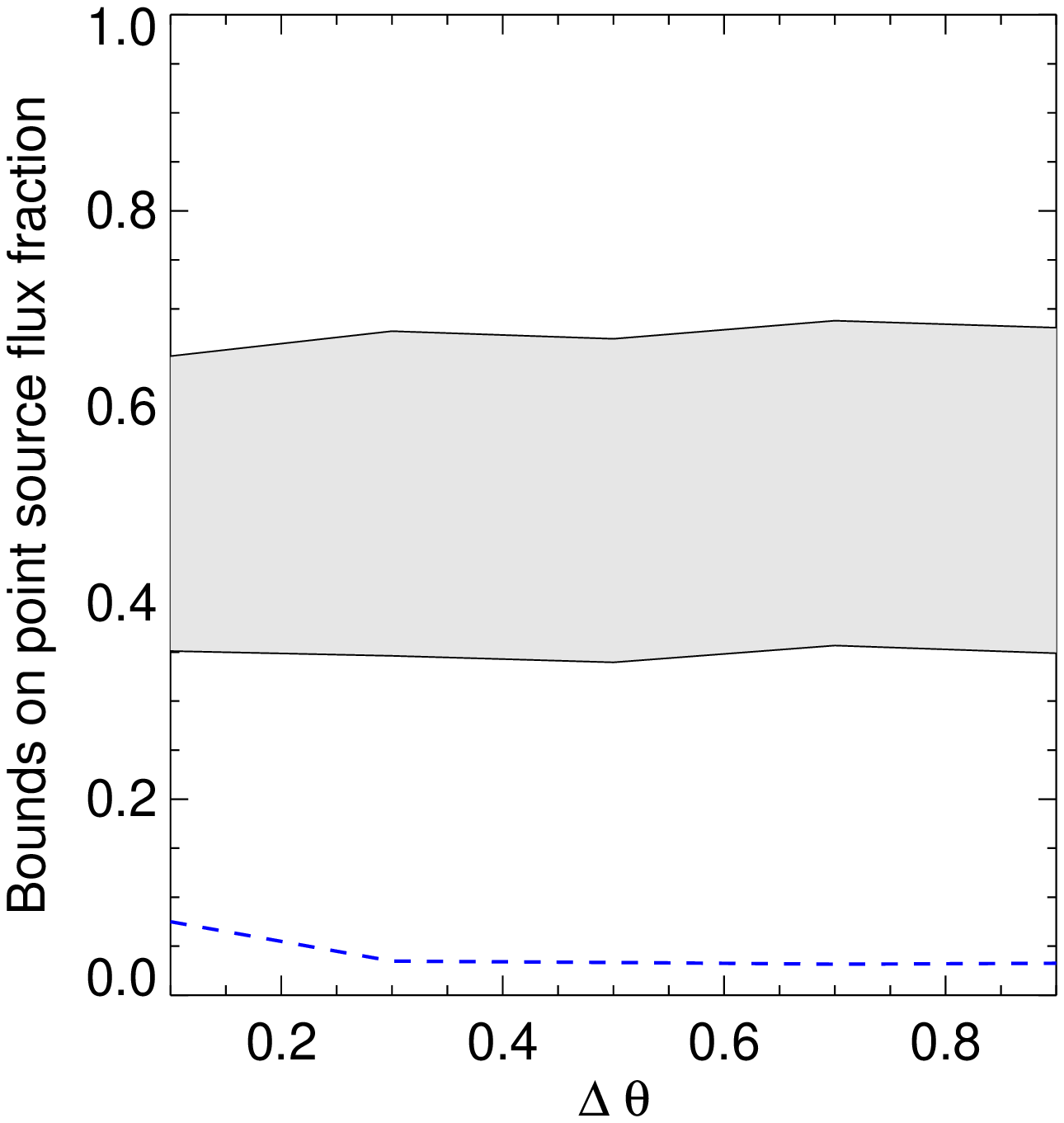}
\includegraphics[width=0.3\textwidth]{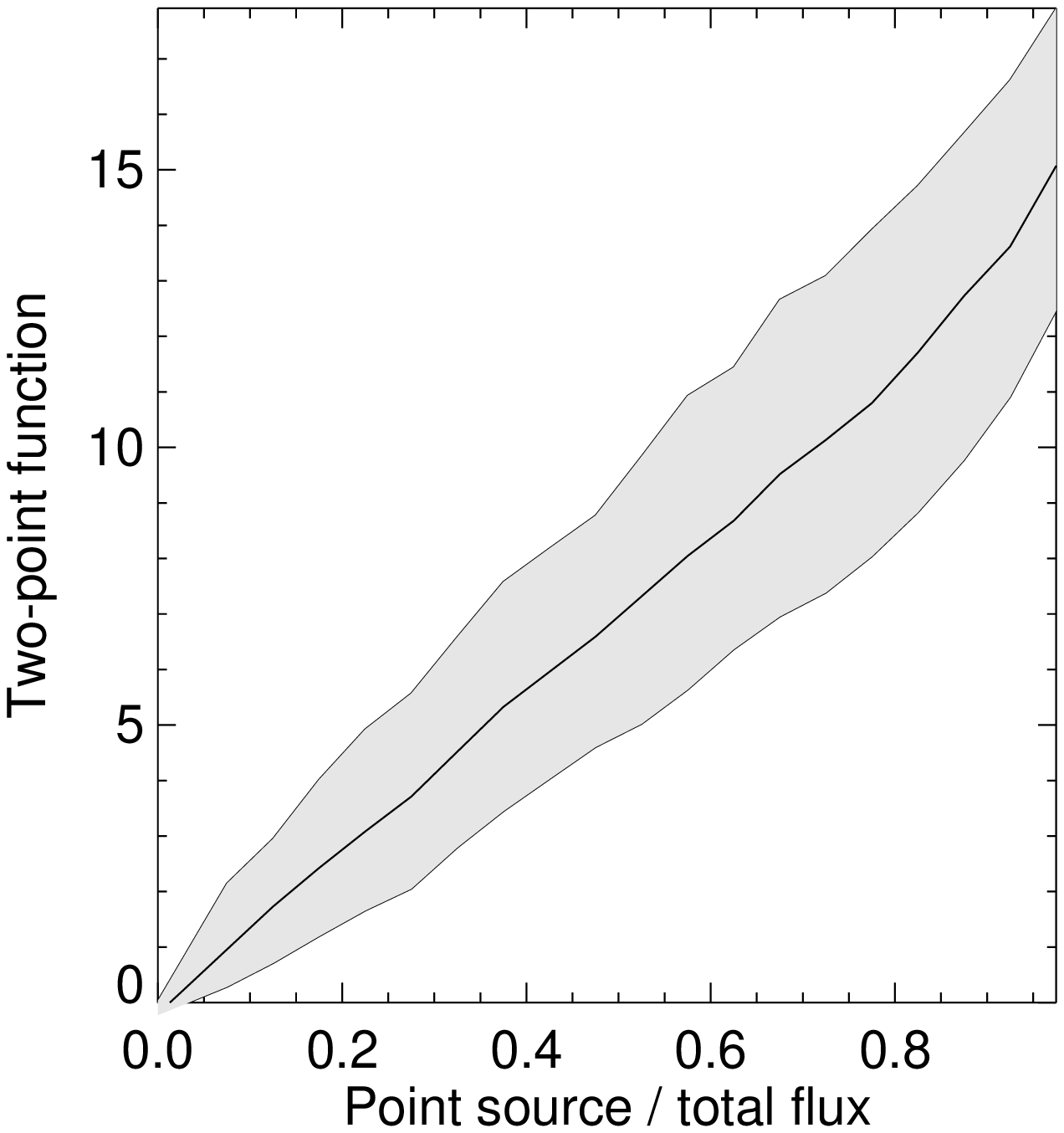}
\caption{\label{fig:twopointfunction} 
Monte Carlo realizations of the two-point correlation function for the angular region and number of events described in \S \ref{sec:mc}, varying the annulus width as a fraction of the test radius (the fractional annulus width is denoted $\Delta \theta$) and taking $r$=$0.2^\circ$=$1 \sigma$ of the PSF, for (\emph{left}) benchmark model 1, and (\emph{center}) benchmark model 2. The right-hand panel shows an example of the mean and 5$\%$ and 95 $\%$ quantiles for the two-point function as a function of the point source flux fraction, in benchmark model 2 with $\Delta \theta = 0.5$.}
\end{figure*}

We can directly compare these two statistics as measures of the point source flux. We employ the unbiased estimator for the two-point function described by \citet{1993ApJ...412...64L}, and compute its value as a function of the point source flux contribution (we also scan over the annulus width $\Delta \theta$), within the Monte Carlo framework described previously. Applying the sensitivity measures previously described, the results for the two benchmark parameter sets are shown in Fig. \ref{fig:twopointfunction}. As expected from the discussion above, we see that in the benchmark 1 case, the results are very similar to those previously found, whereas in the more ``optimistic'' benchmark 2 case, the results are similar in the diffuse limit, but the $R$ test provides much better bounds when the true point source flux fraction is $0.5$.

\section{An example application: high energy \Fermi LAT data from the Haze region}

The \Fermi Gamma-ray Space Telescope has recently released all-sky photon data from its first year of operation \footnote{See \texttt{http://fermi.gsfc.nasa.gov/ssc/data/}}. The diffuse emission measured by \Fermi may include signatures of new physics, such as dark matter annihilation or decay, or photons from new classes of astrophysical sources. In investigating the origin of the diffuse gamma rays, it will be necessary to estimate what fraction of the observed flux could be due to unresolved point sources. The statistic presented here is well suited to address this question, especially at high energy where the PSF of the \Fermi LAT is small and the count rate is low.

A microwave excess termed the ``WMAP Haze'' has been observed in the inner $25^\circ$ of the Galaxy, and attributed to synchrotron radiation from some new population of 10-1000 GeV electrons (and/or positrons) \citep{Dobler:2008ww}. Recent cosmic ray experiments have also measured a rise in the positron fraction at $10-100$ GeV \citep{Adriani:2008zr}, and a hardening in the $e^+ + e^-$ spectrum at 300 GeV - several TeV \citep{aticlatest, Abdo:2009zk}, consistent with a new source of high energy $e^+ e^-$. The \Fermi LAT can constrain any such new source of electrons by searching for gamma rays from inverse Compton scattering of the electrons on starlight.

As an example of how this statistic can be applied, we consider the Class 3 (diffuse class) events measured by the \Fermi LAT with energies between 10-100 GeV, in the region of the sky optimized for study of the WMAP Haze, defined in Galactic coordinates by $|l| < 15, \, -30 < b < -10$. There are 1146 such events in this signal window. We take the signal region $|l| < 18, \, -40 < b < -8$. Above 10 GeV the $68 \%$ containment radius of the LAT is $\lesssim 0.2^\circ$ \citep{Rando:2009yq}, which corresponds to a PSF of $\sim 0.13^\circ$ in the sense that we have used (1D $\sigma$ for the Gaussian distribution of photons from a single point source).  We consider three values of the PSF: the best-estimate upper bound of $0.13^\circ$, and also $0.1^\circ$ and $0.2^\circ$, to demonstrate the possible effect of uncertainties in the PSF, or a varying PSF over the energy range of interest. In all cases we take the test radius $r$ to be equal to the PSF, and generate $10^6$ random points to determine $n_E$.

\begin{figure*}
\includegraphics[width=0.3\textwidth]{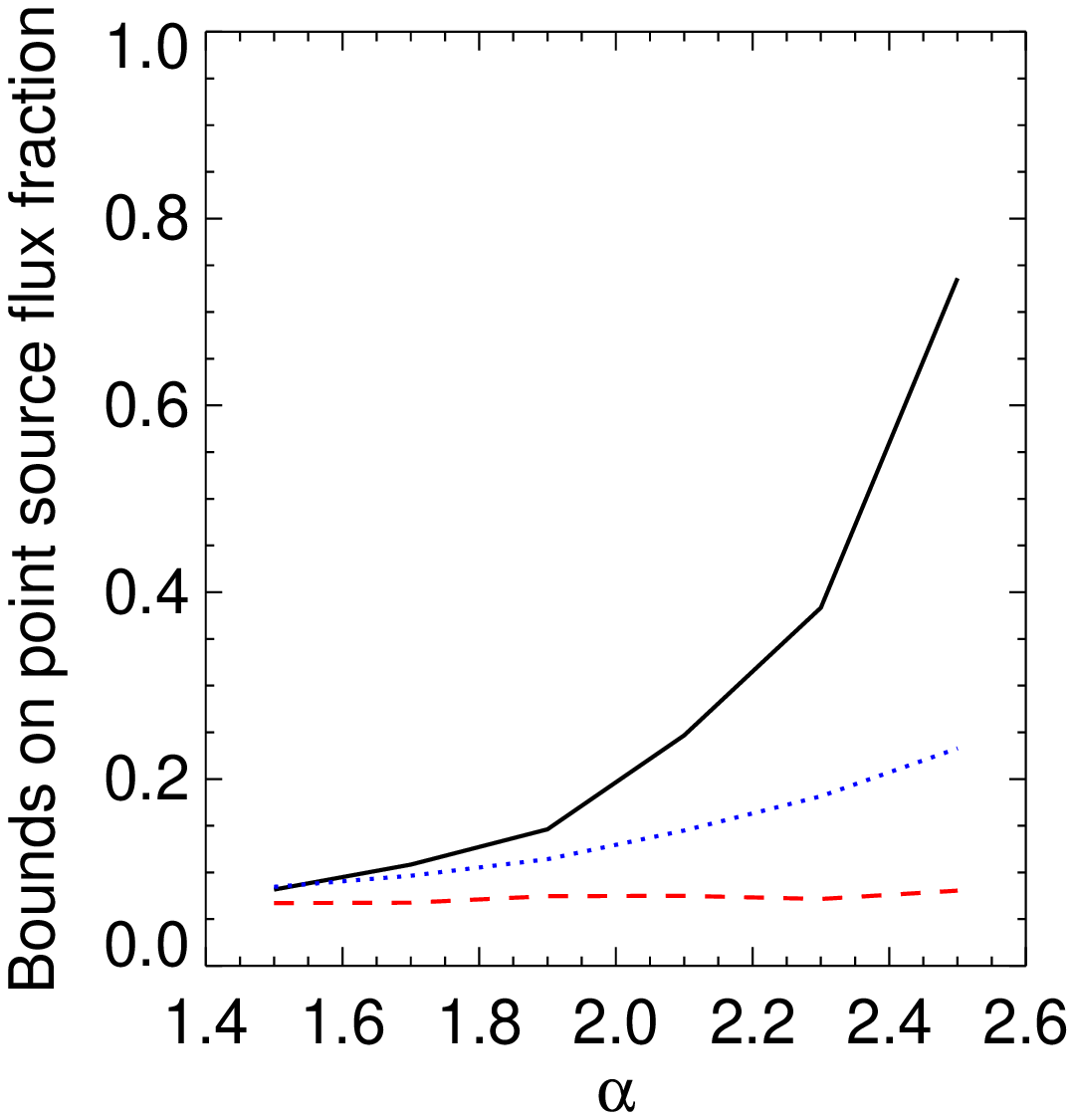}
\includegraphics[width=0.3\textwidth]{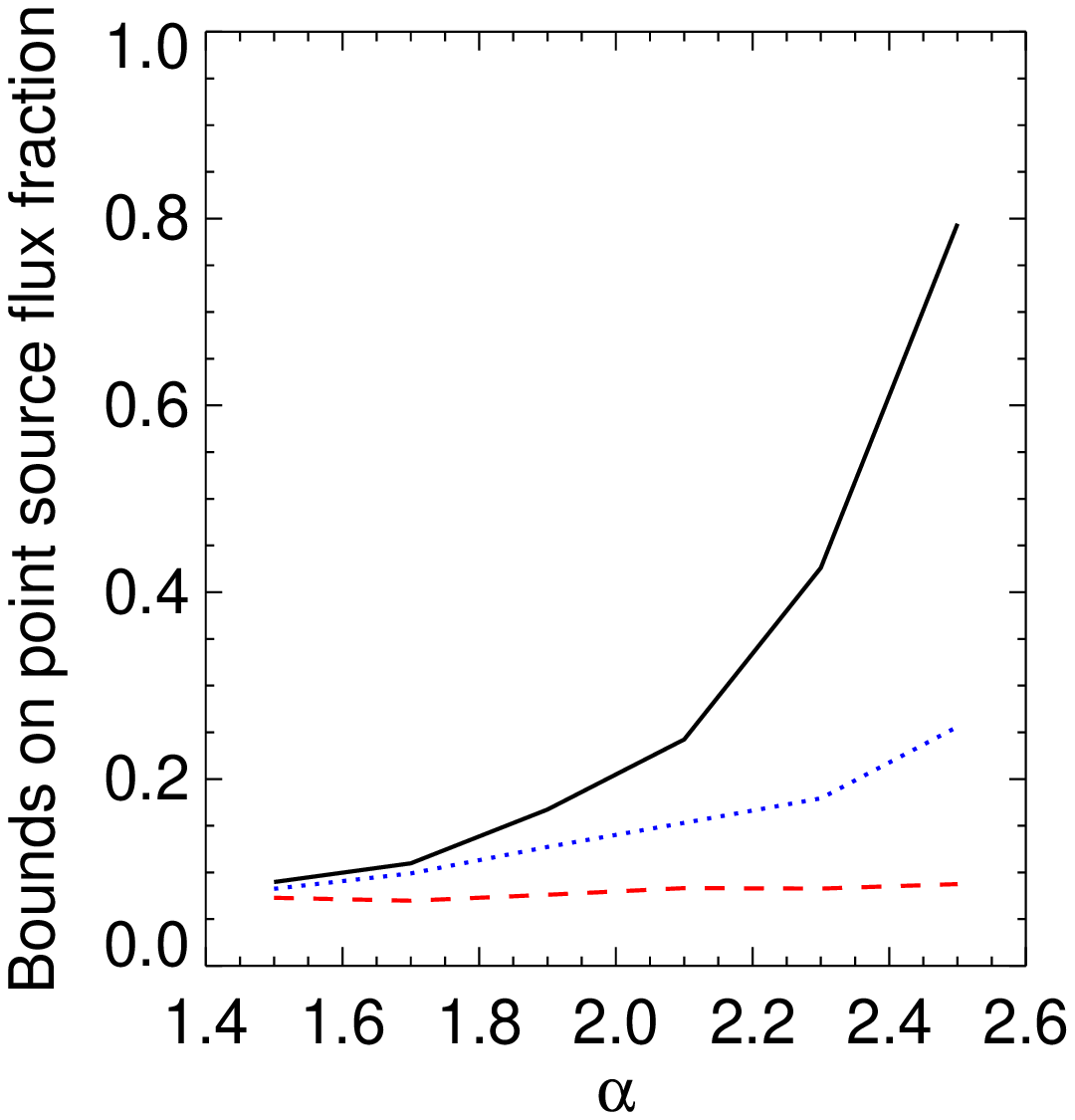}
\includegraphics[width=0.3\textwidth]{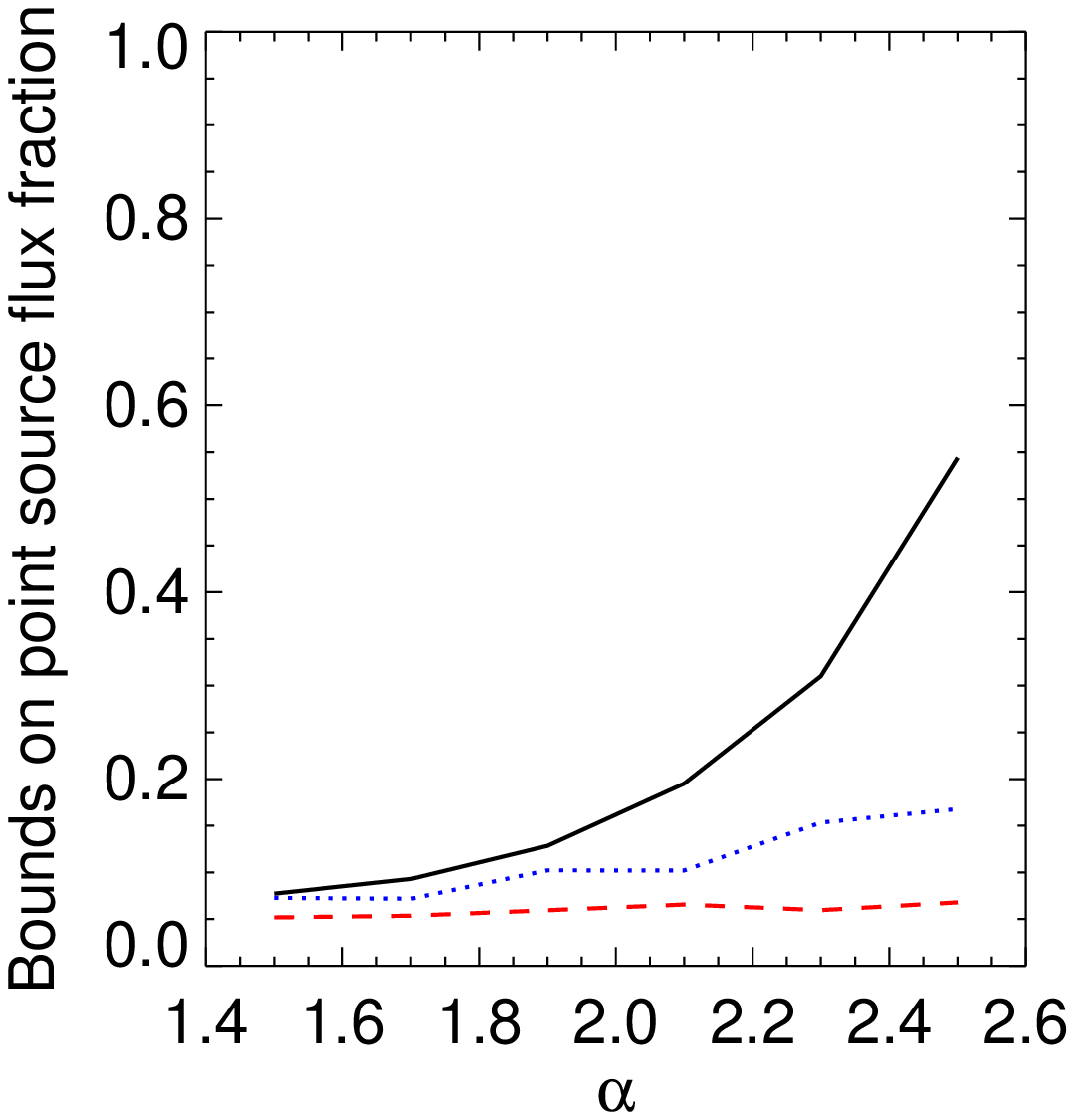}
\caption{\label{fig:fermibenchmarks} 
Bounds on the point source flux fraction in \Fermi LAT data in the ``Haze'' region at 10-100 GeV, from the isotropy ratio $R$ in Monte Carlo simulated data, as a function of the spectral index of the luminosity function. In all cases $S_\mathrm{max} = 10$. \emph{Left:} $r$ = 1 $\sigma$ PSF $= 0.1^\circ$, \emph{center:} $r$ = 1 $\sigma$ PSF $= 0.13^\circ$ (the estimated PSF for back-converting events at 10 GeV), \emph{right:} $r$ = 1 $\sigma$ PSF $= 0.2^\circ$. \emph{Black solid:} $S_\mathrm{min} = 0.01$, \emph{blue dotted:} $S_\mathrm{min} = 0.1$, \emph{red dashed:} $S_\mathrm{min} = 1$.}
\end{figure*}

Fig. \ref{fig:fermibenchmarks} shows the $95 \%$ confidence limits on the fraction of flux originating from point sources, as a function of $S_\mathrm{min}$ and the luminosity function spectral index $\alpha$. We see that even with fairly pessimistic assumptions for the luminosity function, as in the ``Benchmark 1'' case, the measured values of $R$ in the Fermi data are outside the $95 \%$ confidence limits if the flux from sources with $> 0.1$ (expected) counts /year exceeds $\sim 15 \%$ of the total; the limits can be significantly stronger if the assumed point source luminosity function is shallower or extends to higher flux. Note also that we have \emph{not} subtracted resolved point sources in this region; there are no known point sources in this region of the sky in the \emph{Fermi} three-month bright source list \footnote{\texttt{http://fermi.gsfc.nasa.gov/ssc/data/access/lat/bright\_src\_list/}}.

\section{Conclusion}

We have introduced a simple and easily calculable statistic that linearly traces the fraction of flux arising from diffuse emission, as opposed to unresolved point sources. The statistic is quite insensitive to even pronounced large-scale anisotropies in the diffuse emission, such as might originate from the proximity of a bright region or angular variation in the detector exposure. The linear response of this statistic to flux originating from point sources, and its smaller variance, make it superior to the two-point correlation function as a tracer of emission from unresolved point sources.

The sensitivity of the statistic to point source emission naturally depends on the luminosity function of the point sources, as a sufficiently steep power law extending to sufficiently small luminosities is strictly indistinguishable from diffuse emission. However, the statistic retains discriminatory power for spectral indices up to $\alpha \sim 3$, with a low-luminosity cutoff corresponding to an average of 0.1 counts, and assuming all point sources with average luminosity $\gtrsim 10$ counts are resolved and removed. Known luminosity functions for astrophysical point sources generically have shallower slopes than this limit.

When the average number of events per PSF circle exceeds 1, the original form of the statistic breaks down: however, we have described a simple generalization suitable for this case, and demonstrated its efficacy. Increasing the number of counts by taking additional sky regions into account (i.e. without increasing the density of events) improves the variance by the usual $1/N_\mathrm{event}$ Poisson factor. This statistic generalizes readily to higher dimensions; possible applications include the study of void statistics \citep{1986ApJ...306..358F}.

As an example, we have applied this statistic to Class 3 (diffuse class) photon data from the \Fermi LAT in the angular region relevant for study of the WMAP Haze, at energies of 10-100 GeV. We find that even with rather pessimistic assumptions for the point source luminosity function, at most $\sim 15 \%$ of the emission in this region can be attributed to unresolved point sources with average luminosities of $0.1+$ counts / year, and the results are consistent with $100 \%$ diffuse emission.

We wish to acknowledge helpful conversations with Marc Davis, Josh Grindlay, Igor Moskalenko, Jim Peebles and Pat Slane. We thank the anonymous referee for helpful comments. TRS is supported by a Sir Keith Murdoch Fellowship from the American Australian Association. 

\begin{appendix}

\begin{onecolumn}

\section{The analytic grid model and the luminosity function}
\label{app:lfanalytic}

The analytic estimates for $R$ and $\sigma(R)$ derived in \S \ref{sec:analytic} are functions of the fraction of cells which gain exactly 0-1 counts from point sources. If we are to compare the analytic estimates to the results of the MC runs, the $S$ and $T$ parameters must be expressed in terms of $\alpha$, $S_\mathrm{min}$ and $S_\mathrm{max}$. This is a nontrivial exercise, and in any case the analytic estimates and the MC runs should not be expected to agree in detail, since they use different criteria for determining neighboring events (events within the same cell vs events within the test radius). The main purpose of the analytic estimates is to demonstrate the general scaling behavior of $R$ and $\sigma(R)$, and the regions of parameter space where this test loses discriminatory power.

Nonetheless, for completeness, we now derive approximate relations between $S$ and $T$ and the parameters of the luminosity function. These relations are used to provide an analytic estimate for the sensitivity of $R$ as a measure of the point source flux fraction, in Figs. \ref{fig:sdependence}-\ref{fig:varyingr}. Here we only derive the \emph{mean} values for $S$ and $T$ for a given luminosity function; for any choice of the luminosity function, there will be an additional contribution to $\sigma^2(R)$ from the variances of $S$ and $T$, which is not taken into account in this analysis.

Let us first compute the probabilities for a given $(i,j)$ cell to obtain exactly one or zero counts from the addition of a single (randomly placed) point source, denoted $\rho_1^{ij}$ and $\rho_0^{ij}$ respectively. If the point source has an expected contribution of $k$ counts, the PSF is assumed to be Gaussian with variance $\sigma^2$, the cells are square with dimensions $\delta \times \delta$, and the point source is centered at $(x_0, y_0)$, then the expected number of counts in a cell with left-hand lower corner $(x_i, y_j)$ is given by,
\begin{eqnarray} \lambda_{i j}(k,x_0,y_0) & = & \frac{k}{2 \pi \sigma^2} \int^{x_i+\delta}_{x_i} \int^{y_j + \delta}_{y_j} e^{-(x-x_0)^2/2 \sigma^2} e^{-(y-y_0)^2/2 \sigma^2} dx dy, \nonumber \\ & = & \frac{k}{4} \left(\mathrm{Erf}\left( \frac{x_0 - x_i}{\sqrt{2} \sigma} \right) - \mathrm{Erf}\left( \frac{x_0 - x_i - \delta}{\sqrt{2} \sigma} \right) \right) \nonumber \\ & & \times \left(\mathrm{Erf}\left( \frac{y_0 - y_j}{\sqrt{2} \sigma} \right) - \mathrm{Erf}\left( \frac{y_0 - y_j - \delta}{\sqrt{2} \sigma} \right) \right). \label{eq:lambda} \end{eqnarray}

The Poisson probabilities to obtain exactly one and zero counts in this cell, from a point source providing $k$ events, are then given by,
\begin{equation} \rho_1^{ij}(k,x_0,y_0) = \lambda_{ij}(k,x_0,y_0) e^{-\lambda_{ij}(k,x_0,y_0)}, \quad \rho_0^{ij}(k,x_0,y_0) = e^{-\lambda_{ij}(k,x_0,y_0)}. \end{equation}
Since the $\lambda_{ij}$'s depend only linearly on $k$, it is straightforward to integrate over the luminosity function: writing $\lambda_{ij} = k \theta$, we obtain,
\begin{eqnarray} \rho_0^{ij}(x_0, y_0) & = & \frac{\int^{S_\mathrm{max}}_{S_\mathrm{min}} k^{-\alpha} e^{-k \theta} dk}{ \int^{S_\mathrm{max}}_{S_\mathrm{min}} k^{-\alpha} dk} \nonumber \\ & = & \frac{\theta^{\alpha - 1} \left( \Gamma\left( 1 - \alpha, S_\mathrm{min} \theta \right) - \Gamma\left( 1 - \alpha, S_\mathrm{max} \theta \right) \right) (1 - \alpha)}{S_\mathrm{max}^{1 - \alpha} - S_\mathrm{min}^{1 - \alpha}}, \end{eqnarray}
\begin{eqnarray} \rho_1^{ij}(x_0, y_0) & = & \frac{\int^{S_\mathrm{max}}_{S_\mathrm{min}} k^{-\alpha} k \theta e^{-k \theta} dk}{ \int^{S_\mathrm{max}}_{S_\mathrm{min}} k^{-\alpha} dk} \nonumber \\ & = & \frac{\theta^{\alpha - 1} \left( \Gamma\left( 2 - \alpha, S_\mathrm{min} \theta \right) - \Gamma\left( 2 - \alpha, S_\mathrm{max} \theta \right) \right) (1 - \alpha)}{S_\mathrm{max}^{1 - \alpha} - S_\mathrm{min}^{1 - \alpha}}. \label{eq:probs} \end{eqnarray}

Strictly speaking we should now integrate this result with respect to $x_0$ and $y_0$; however, this is not analytically tractable. Provided $\sigma$ is not too much smaller than $\delta$, it is a good approximation to instead integrate Eq. \ref{eq:lambda} over $x_0$ and $y_0$ within a given cell, obtaining an average expected number of counts $\lambda_{ij}(k) = k \theta$ for each cell, and then use this result for $\theta$ in Eq. \ref{eq:probs}. Let us choose our coordinate system so that $(x_0, y_0)$ lies in a cell with left-hand corner $(0,0)$, and the $i j$ cell has left-hand corner $(i \delta, j \delta)$. For $|i|$ or $|j| \gg 0$, $\lambda_{ij}$ will be negligible, so we can make the further approximation of truncating the sum over $i, j$ at some point. In this work we make the approximation that for cells with $|i|$ or $|j| > 2$, $\rho_0^{ij} = 1$ and $\rho_1^{ij} = 0$.

We can then write $\theta$ for each cell in terms of three functions of $\sigma/\delta$, denoted $t$, $u$ and $v$, obtained by averaging $(1/2) \left(\mathrm{Erf}\left( \frac{x_0 - x_i}{\sqrt{2} \sigma} \right) - \mathrm{Erf}\left( \frac{x_0 - x_i - \delta}{\sqrt{2} \sigma} \right) \right)$ over $x_0 = [0, \delta]$, for $i=0, 1, 2$ respectively:
\begin{eqnarray} t & = & \mathrm{Erf} \left(\frac{\delta}{\sqrt{2} \sigma} \right) - \sqrt{\frac{2}{\pi}} \frac{\sigma}{\delta} \left(1 - e^{-\delta^2 / 2 \sigma^2} \right) , \nonumber \\
 u & = & \frac{\sigma}{\delta \sqrt{2 \pi}} \left(e^{-\frac{2 \delta^2}{\sigma^2}} -2 e^{\frac{ \delta^2}{2 \sigma^2}}+1\right) - \text{Erf}\left(\frac{\delta}{\sqrt{2} \sigma}\right)+ \text{Erf}\left(\frac{\sqrt{2} \delta}{\sigma}\right), \nonumber \\
 v & = & \frac{1}{2} \left(\text{Erf}\left(\frac{\delta}{\sqrt{2} \sigma}\right)+3 \text{Erf}\left(\frac{3 \delta}{\sqrt{2} \sigma}\right)-4 \text{Erf}\left(\frac{\sqrt{2} \delta}{\sigma}\right)\right)  + \frac{\sigma}{\delta \sqrt{2 \pi }}\left(e^{-\frac{9 \delta^2}{2 \sigma^2}} -2 e^{-2 \delta^2 / \sigma^2} +e^{- \delta^2 / 2 \sigma^2} \right). \end{eqnarray}

In terms of these functions the values of $\theta$ for the relevant cells are:
\[i=j=0, \quad 1 \, \, \mathrm{cell}, \quad \theta = t^2 \]
\[i=0, \, j= \pm 1 \quad \mathrm{or} \quad i=\pm 1, \, j = 0, \quad 4 \, \, \mathrm{cells}, \quad \theta = t u \]
\[i=\pm 1, \, j= \pm 1, \quad 4 \, \, \mathrm{cells}, \quad \theta = u^2 \]
\[i=0, \, j= \pm 2 \quad \mathrm{or} \quad i=\pm 2, \, j = 0, \quad 4 \, \, \mathrm{cells}, \quad \theta = t v \]
\[i=\pm 1, \, j= \pm 2 \quad \mathrm{or} \quad i=\pm 2, \, j = \pm 1, \quad 8 \, \, \mathrm{cells}, \quad \theta = u v \]
\[i=\pm 2, \, j= \pm 2, \quad 4 \, \, \mathrm{cells}, \quad \theta = v^2. \]

Summing over the probabilities $\rho_0^{ij}$ ($\rho_1^{ij}$) for all cells then yields the expected number of cells which gain zero (one) counts from the addition of a point source, denoted $E_0$ ($E_1$). If $n$ sources are added, the probability of any one cell gaining zero counts is $(E_0/N)^n$, and the probability of a single count is $n (E_0/N)^{n-1} (E_1/N)$ (i.e. one source contributes a single count, all others contribute zero). Multiplying by the number of cells $N$ yields the $N - T$ and $S$ parameters respectively. The number of sources $n$ is related to the average total emission from point sources,
\begin{equation} n = \frac{2 - \alpha}{1 - \alpha} \left( \frac{S_\mathrm{max}^{1 - \alpha} - S_\mathrm{min}^{1-\alpha}}{S_\mathrm{max}^{2 - \alpha} - S_\mathrm{min}^{2-\alpha}} \right) \times \mathrm{mean} \, \mathrm{total} \, \mathrm{counts}. \end{equation}

In order to compare the analytic result (based on the grid) to the MC results (based on neighbors within a test radius), we must also impose a relation between the side length of the grid cells and the test radius. In Figs. \ref{fig:sdependence}-\ref{fig:varyingr} we require that the area of a grid cell is the same as the area of a PSF circle, i.e. $\delta = \sqrt{\pi} r$; a different prescription might give better agreement between the MC results and the estimates from the grid model.

\end{onecolumn}

\end{appendix}

\bibliographystyle{mn2e_eprint}
\bibliography{fsratio}

\end{document}